\documentclass[structabstract]{aa}

\usepackage{amsmath}
\usepackage{txfonts}
\usepackage{booktabs}
\usepackage{array}
\usepackage{graphicx}
\usepackage{natbib,twoopt}
\usepackage{upgreek}
\usepackage{color}
\usepackage[breaklinks=true]{hyperref} 
\bibpunct{(}{)}{;}{a}{}{,} 
\newcommandtwoopt{\citeads}[3][][]{\href{http://adsabs.harvard.edu/abs/#3}%
{\citealp[#1][#2]{#3}}}
\newcommandtwoopt{\citepads}[3][][]{\href{http://adsabs.harvard.edu/abs/#3}%
{\citep[#1][#2]{#3}}}
\newcommandtwoopt{\citetads}[3][][]{\href{http://adsabs.harvard.edu/abs/#3}%
{\citet[#1][#2]{#3}}}
\newcommandtwoopt{\citeyearads}[3][][]%
{\href{http://adsabs.harvard.edu/abs/#3}{\citeyear[#1][#2]{#3}}}

\begin{document}

\title{\object{M82} - A radio continuum and polarisation study}
\subtitle{I. Data reduction and cosmic ray propagation}
\author{B. Adebahr\inst{1}\and M. Krause\inst{2} \and U. Klein\inst{3} \and M. We\.zgowiec\inst{1} \and D. J. Bomans\inst{1} \and R.-J. Dettmar\inst{1}}
\institute{Astronomisches Institut der Ruhr-Universit\"at Bochum (AIRUB), Universit\"atsstrasse 150, 44801 Bochum, Germany
\and Max-Planck-Institut f\"ur Radioastronomie (MPIfR), Auf dem H\"ugel 69, 53121 Bonn, Germany
\and Argelander-Institut f\"ur Astronomie (AIfA), Universit\"at Bonn, Auf dem H\"ugel 71, 53121 Bonn, Germany}
\date{Received 14 August 2012 / Accepted -}
\abstract {The potential role of magnetic fields and cosmic ray propagation for feedback processes in the early Universe can be probed by studies of local starburst counterparts with an equivalent star-formation rate.} {In order to study the cosmic ray propagation, determine the magnetic field strength and the dominant loss processes in the nearby prototypical starbursting galaxy M82 a multi-frequency analysis at four radio wavelengths is presented.} {Archival data from the WSRT was reduced and a new calibration technique introduced to reach the high dynamic ranges needed for the complex source morphology. This data was combined with archival VLA data, yielding total power maps at $\uplambda$3\,cm, $\uplambda$6\,cm, $\uplambda$22\,cm and $\uplambda$92\,cm.} {The data shows a confinement of the emission at wavelengths of $\uplambda$3/$\uplambda$6\,cm to the core region and a largely extended halo reaching up to 4\,kpc away from the galaxy midplane at wavelengths of $\uplambda$22/$\uplambda$92\,cm up to a sensitivity limit of $90\,\upmu$Jy and $1.8\,$mJy respectively indicating different physical processes in the core and halo regions. The results are used to calculate the magnetic field strength in the core region to $98\,\upmu$G and to $24\,\upmu$G in the halo regions. From the observation of ionisation losses the filling factor of the ionised medium could be estimated to 2\,\%. This leads to a revised view of the magnetic field distribution in the core region and the propagation processes from the core into the halo regions.} {We find that the radio emission from the core region is dominated by very dense \ion{H}{ii}-regions and supernova remnants, while the surrounding medium is filled with hot X-ray and neutral gas. Cosmic rays radiating at frequencies higher than 1.4\,GHz are suffering from high synchrotron and inverse Compton losses in the core region and are not able to reach the halo. Even the cosmic rays radiating at longer wavelengths are only able to build up the observed kpc sized halo, when several starbursting periods are assumed where the FIR- and radio luminosity vary by an order of magnitude. These findings together with the strong correlation between H$\upalpha$, PAH$^+$, and our radio continuum data suggests a magnetic field which is frozen into the ionised medium and driven out of the galaxy kinematically.}


\maketitle

\section{Introduction}

M82 is a nearby prototypical starburst galaxy with a distance of $\sim$3.5\,Mpc \citepads{2009AJ....138..332J} in the M81 group of galaxies. It hosts a complicated distribution of different gas phases in its halo as well as in its violently starbursting core. Material from the core region is accelerated into the halo via a galactic superwind as well as dragged out of the galaxy by tidal interaction. A prominent large scale biconical outflow has been observed in H$\upalpha$ \citepads{1999ApJ...510..197D,2002PASJ...54..891O} as well as in X-rays \citepads{2009ApJ...697.2030S}. Dust, neutral gas phases and even molecular material are mixed in the ionised phase and seem to be swept along by this superwind (e.g. CO: \citetads{2002ApJ...580L..21W}; H$_2$: \citetads{2009ApJ...700L.149V}; Dust: \citetads{2002PASJ...54..891O,2009AJ....137..517L}). Large-scale molecular and atomic streamers are anchored at the edges of the galactic disk and therefore show a completely different distribution \citepads{2002ApJ...580L..21W,1993ApJ...411L..17Y}. These are interpreted to be remnants of the interaction with M81 approximately 100\,Myrs ago \citepads{1993egte.conf..253Y}.

While these kinematic parameters are confirmed by a large number of observations and studies, the contribution of the magnetic field to the outflow is still a matter of debate. For normal spiral galaxies the mean field $\upalpha\upomega$-dynamo in combination with galactic winds driven by supernova explosions in the spiral arms is the favoured explanation for the observed magnetic field patterns \citepads{1996ARA&A..34..155B,2009RMxAC..36...25K}. However, the role of the magnetic field in starburst galaxies is completely unclear. It can be a passive one, where the magnetic field is simply dragged along the velocity gradient of the wind, or an active one, where the magnetic field influences the gas motions in the outflow and it becomes a wind driven by cosmic rays and its associated field. The total magnetic field energy in normal spiral galaxies contributes only marginally to the energy budget, but the high particle densities and possible high magnetic field strengths in starbursts can increase it to a value where it needs to be accounted for. A detailed study of these parameters is needed to constrain the inputs for simulations to understand material circulation and the observed morphology of galactic halos. This has a significant influence on models for the origin of magnetic fields and the magnetisation of the intergalactic medium in the early Universe, where starbursts and interactions are known to be more common. Starbursting dwarf galaxies in the early Universe can explain the magnetisation and enrichment of the intergalactic medium without a primordial seed field \citepads{1999ApJ...511...56K}.

Current instruments lack the sensitivity as well as the resolution for a study of such objects in the early Universe, so that an extrapolation from nearby observable targets is needed. Due to its proximity and large apparent size of 11\farcm2$\times$4\farcm3 and the available data at nearly all wavelengths, M82 is the ideal target for such an analysis. In addition, its interaction scenario makes it possible to examine the influence of the group medium on a galactic superwind scenario hosting a starburst.

\citetads{1994Natur.372..530Y} and \citetads{2008AJ....135.1983C} found a large amount of \ion{H}{i} filling the whole group as well as three distinct spurs, which connect the main galaxies of the group: M81, M82, and NGC3077. Additionally, M82 is well known for its large fraction of molecular gas in the central region \citepads{2011A&A...535A..84A,2010ApJ...722..668N}. Observations by \citetads{2000Natur.404..732G} show a giant magnetic molecular bubble possibly blown out by the superwind. But how far this gas distribution influences the evolution of the whole group and whether it is reaccreted to their host galaxies is still not clear. An extensive star formation in the recently formed spurs could have enriched the intragroup medium with metals and magnetic fields \citepads{2008AJ....135..548D}.

In how far the superwind in M82 is constantly fuelling the halo with fresh material is a matter of discussion. Evolution scenarios reach from an increased star-formation period lasting more than 10\,Myrs to the occurrence of several episodes of increased star-formation over the last Myrs. A detailed NIR-MIR spectroscopy by \citetads{2003ApJ...599..193F} suggests two successive starburst periods about 10 and 5\,Myrs ago. This would explain the enrichment of the halo with ionised polycyclic aromatic hydrocarbons (PAH) observed by \citetads{2010A&A...514A..14K}, which would otherwise have been destroyed by the starburst of the violent radiation. Additionally extended emission has been found $\sim$11\,kpc to the north of the galaxy in H$\upalpha$, X-rays, and UV \citepads{1999ApJ...510..197D,1999ApJ...523..575L,2005ApJ...619L..99H}, which is spatially not connected to the main halo of M82. This might be a collision of the hot superwind with a preexisting neutral cloud \citepads{1999ApJ...523..575L} and a remnant of a previously stronger star-bursting activity blowing material even further into the halo. The detection of hard $\upgamma$-ray emission by the \citetads{2009Natur.462..770V} further verifies a closed environment scenario since it favours the idea of M82 being a proton calorimeter \citepads{2011ApJ...734..107L}. Additionally the high particle and energy densities in the core region lead to a scenario where even charged particles accelerated by supernova explosions would not be able to escape this region and could not build up a synchrotron halo assuming the recent activity of star-formation and particle densities.

Previous studies by \citetads{1991ApJ...369..320S} and \citetads{1992A&A...256...10R} found emission, which was mostly confined to the core region of the galaxy and discussed the possibilities for reacceleration and/or dynamo processes \citepads{1994A&A...282..724R} inside the turbulent medium of M82, but could not sufficiently explain the diffuse emission reaching up into the halo. The improved data reduction and instrumental techniques together with more precise measurements over the whole wavelengths regime now allows us to study this prototypical starburst galaxy with increased detail.

Since the synchrotron emission from gyrating electrons is the best tracer for magnetic fields, a complementary study at several different radio wavelengths between $\uplambda3$\,cm and $\uplambda92$\,cm is needed to get a full picture of the loss processes and the magnetic field morphology of this galaxy and its surrounding halo. The high flux difference between the highly luminous core and the very diffuse halo structure especially at longer wavelengths makes such a study challenging for the dynamic range requirements of today's available instruments and data reduction techniques. Therefore data from the VLA at $\uplambda3$\,cm and $\uplambda6$\,cm and data from the WSRT at $\uplambda22$\,cm and $\uplambda92$\,cm were used for the following study. Both telescopes are well known for their exceptionally good properties at these mentioned wavelengths.

The goal of this paper is to investigate the outflow mechanism and to describe a revised scenario for the built-up of the observed radio halo. As a first step we investigate the magnetic field strength in M82 using an energy equipartition ansatz. These results can then be used to calculate the different loss timescales and designate the dominant physical processes influencing the wind in the two different regions. Arguments for the existing halo will be discussed and a scenario with a low filling factor of the ionised as well as magnetised medium in the core region is described. These results and the comparison with observations at other wavelengths will be used to understand the outflow mechanism of this archetypal galaxy and draw conclusions for the early Universe.

This paper is organised as follows: Section \ref{text_datareduction} shows the reduced data and explains the special data reduction techniques used to reach the needed high dynamic ranges for this study. In Section \ref{text_totalpower} the total intensity maps and vertical scaleheights are presented, Section \ref{text_spectralindex} shows the spectral index maps. Estimates for the magnetic field strength are given in Section \ref{text_magneticfield} and a detailed discussion about the different cosmic ray losses is presented in Section \ref{label_losses}. We summarise our outcomes in Section \ref{text_summary} and discuss them in Section \ref{text_conclusions}.

\begin{table}
	\caption{Basic properties of M82}
	\centering
	\begin{tabular}{@{} lc @{}}
		\toprule
		\toprule
		Name & M82 \\
		Alternative Names & NGC3034, 3C231, Arp337 \\
		RA$_{2000}$ & 09\fh55\fm52\fs7 \\
		DEC$_{2000}$ & +69\degr40\arcmin46\arcsec \\
		Type & I0; Starburst$^\text{a}$ \\
		Apparent Size (D$_{25}$) & $11\farcm2\times 4\farcm3$$^\text{d}$\\
		Distance & 3.52\,Mpc$^\text{b}$ \\
		Parallactic angle & 68\degr$^\text{c}$ \\
		Inclination & 79\degr$^\text{c}$ \\
		\bottomrule
	\end{tabular}
	\begin{list}{}{}
	\item[]$^\text{a}$NED $^\text{b}$\citetads{2009AJ....138..332J} $^\text{c}$HyperLeda $^\text{d}$\citetads{1998A&AS..127..269H}
	\end{list}
\end{table}

\section{Observations and data reduction}
\label{text_datareduction}

\begin{table*}
	\caption{Observation parameters}
	\label{table_obs}
	\centering
	\newcolumntype{1}{>{\centering\arraybackslash} m{1.4cm} }
	\newcolumntype{2}{>{\centering\arraybackslash} m{1.3cm} }
	\newcolumntype{3}{>{\centering\arraybackslash} m{1.4cm} }
	\newcolumntype{4}{>{\centering\arraybackslash} m{1.4cm} }
	\newcolumntype{5}{>{\centering\arraybackslash} m{1.7cm} }
	\newcolumntype{6}{>{\centering\arraybackslash} m{1.5cm} }
	\newcolumntype{8}{>{\centering\arraybackslash} m{1.6cm} }
	\newcolumntype{9}{>{\centering\arraybackslash} m{1.5cm} }
	\newcolumntype{0}{>{\centering\arraybackslash} m{1.5cm} }
	\renewcommand{\arraystretch}{1.5}
	\begin{tabular}{@{} 1 2 3 0 4 5 6 8 9 @{}}
		\toprule
		\toprule
		Wavelength (cm) & Telescope and configuration & Total bandwidth (MHz) & Channels per subband & Channel bandwidth (MHz) & Subband frequencies (MHz) & Polarisation products & Total inte- gration time (min) & Observing date \\
		\midrule
		3 & VLA-D & 100 & 1 & 50 & 8465, 8415 & RR, LL, RL, LR & 160 & 1990/01/04 \\
		6 & VLA-D & 100 & 1 & 50 & 4885, 4835 & RR, LL, RL, LR & 160 & 1990/01/04 \\
		6 & VLA-D & 100 & 1 & 50 & 4885, 4835 & RR, LL, RL, LR & 225 & 1986/01/26 \\
		18 & WSRT & 160 & 64 & 0.313 & 1753, 1737, 1721, 1705, 1689, 1673, 1657, 1641 & XX, YY, XY, YX & 360 & 2003/03/16 \\
		22 & WSRT & 160 & 64 & 0.313 & 1422, 1406, 1390, 1374, 1358, 1342, 1326, 1310 & XX, YY, XY, YX & 360 & 2003/03/16 \\
		22 & WSRT & 80 & 128 & 0.078 & 1450, 1424, 1419, 1398, 1374, 1349, 1326, 1303 & XX, YY & 720 & 2003/02/09 \\
		92 & WSRT & 20 & 128 & 0.019 & 321, 325, 332, 336, 372, 376, 380, 385 & XX, YY & 720 & 2003/04/30 \\
		\bottomrule
	\end{tabular}
\end{table*}

\subsection{VLA observations}

The $\uplambda3$\,cm and $\uplambda6$\,cm data were obtained from the archive (already published by \citetads{1991ApJ...369..320S} and \citetads{1992A&A...256...10R}), combined and the calibration improved for our studies. We chose only the data from the compact D-configuration to match the resolution to our WSRT data and to restore most of the diffuse halo structure of M82.

All observations used two subbands at central frequencies of 8.465\,GHz and 8.415\,GHz and 4.885\,GHz and 4.835\,GHz, respectively. Each subband had a bandwidth of 50\,MHz, resulting in a total bandwidth of 100\,MHz for each of the observed wavelengths. The observations were accompanied by calibrator observations of 3C138, 3C286, 3C48, 1044+719, and 0836+710 to determine the fluxes, phase corrections, polarisation leakages and polarisation angles over the time of the observations. Data reduction and flagging was done using the Common Astronomy Software Applications package (CASA). Initially calibrator gains were applied independently to each subband and observation using the flux density scale given by Perley and Butler (2010). Then self-calibration was performed in the same way calibrating four times on phase only followed by one simultaneous amplitude and phase calibration. For each iteration a model was derived using the Clark-Clean algorithm \citepads{1980A&A....89..377C}. Inside each self-calibration loop the clean iterations were increased to include more of the diffuse flux in the model and the time interval for solutions decreased until the integration time was reached. The data were then cleaned down to a $3\sigma$-noise level and baseline-based corrections were applied using a long solution interval of 10 minutes for the total power images to increase the dynamic range. Integrated fluxes for the total power images before and after the baseline-based corrections were applied and differed by less than 1.5\,\%. Not enough flux was available in the Stokes Q and U images, so that the baseline-based corrections were not used for the production of the polarisation images. For each clean step masks were used from previous data reductions iteratively including more sources inside the field. The resulting final images of each observed wavelength were combined using an inverse square weighting followed by primary beam correction.

This subband-based calibration strategy lowers the frequency dependence of the gain solutions as well as the flux variations of sources inside the field over the complete bandwidth. Care has to be taken for observations with low flux intensities since the described method lowers the signal-to-noise for the determination of the gain solutions, which can lead to an image quality worse than calibrating using the whole data in all.

\subsection{Westerbork observations}
\label{text_channel_based}

The WSRT archive was inspected for archival datasets of M82. One dataset at 18\,cm, two at 22\,cm and one at 92\,cm were found. The 18\,cm data and one of the 22\,cm datasets were recorded in all four cross-correlations. Therefore these data contained polarisation information. For the other only the XX- and YY-correlations were available. From this data only a total power image of the 18\,cm dataset has been published so far by \citetads{2007A&A...461..455B}.

All datasets were first read into the Astronomical Image Processing System (AIPS) to apply the system temperatures. Since AIPS can normally only handle circular polarised feeds, the data headers were changed to circular polarisation during this step and changed back to linear polarisation later. The data were converted into the CASA MS-format and inspected using RFI GUI, the graphical frontend to RFI console \citepads{2010MNRAS.405..155O}. A special flagging strategy was developed for each wavelength and source (calibrators/M82). Since the bandpass for the WSRT is very steep at the subband edges, a preliminary bandpass was applied to the data which was not used for calibration during later stages of the data reduction. For this the calibrator observation least affected by RFI (Radio Frequency Interference) was used for the appropriate wavelength. This procedure helped RFI console to distinguish more easily between real RFI and bandpass effects. After this initial flagging each dataset was again checked for leftover RFI, which was then flagged manually.

\begin{table*}
	\caption{Summary of calibrator parameters}
	\label{table_calflux}
	\centering
	\newcolumntype{1}{>{\centering\arraybackslash} m{1.9cm} }
	\newcolumntype{2}{>{\centering\arraybackslash} m{1.1cm} }
	\newcolumntype{3}{>{\centering\arraybackslash} m{1.1cm} }
	\newcolumntype{4}{>{\centering\arraybackslash} m{1.1cm} }
	\newcolumntype{5}{>{\centering\arraybackslash} m{1.1cm} }
	\newcolumntype{6}{>{\centering\arraybackslash} m{1.1cm} }
	\newcolumntype{7}{>{\centering\arraybackslash} m{1.1cm} }
	\newcolumntype{8}{>{\centering\arraybackslash} m{1.1cm} }
	\begin{tabular}{@{} 1 2 3 4 5 6 7 8 @{}}
		\toprule
		\toprule
		Calibrator & $A$ & $B$ & $C$ & $D$ & $RM$ \quad (rad/m$^2$) & $p$ & $\Phi_0$ \quad (rad) \\
		\midrule
		\object{3C138} (22cm) 3C138 (18cm) & 1.0076 & -0.5562 & -0.1113 & -0.0146 & -1.2867 & 0.0750 0.0845 & -0.1379 \\
		\object{CTD93} & 0.7110 & -0.0230 & -1.3190 & 0.4938 & 0 & 0 & 0 \\
		\object{3C147} & 1.4486 & -0.6725 & -0.2112 & 0.0408 & 0 & 0 & 0 \\
		\object{3C295} & 1.4674 & -0.7735 & -0.2591 & 0.0075 & 0 & 0 & 0 \\
		\bottomrule
	\end{tabular}
\end{table*}

Calibrator gains were determined from the unpolarised calibrator, and polarisation leakage and angle terms were determined for the polarised calibrator. This was done for each channel separately. The model parameters for each calibrator were based on a fourth order polynomial with coefficients $A,B,C,$ and $D$ from the VLA Calibrator Manual\footnote{\url{http://www.vla.nrao.edu/astro/calib/manual/baars.html}}. Additionally, the models of the polarised calibrators were given a fractional polarisation $p$, an intrinsic polarisation angle $\Phi_0$ and a Rotation Measure $RM$. This information was gathered from the newest EVLA measurements by R. Perley (priv. comm.). According to the equations
\begin{align}
I & = 10^{A + B\cdot\log{\nu} + C\cdot({\log{\nu})^2} + D\cdot({\log{\nu})^3}}\\
Q & = p \cdot I \cdot \cos\left[{2 \cdot \left( \Phi_0 + RM \cdot \frac{c}{\nu} \right)^2}\right]\\
U & = p \cdot I \cdot \sin\left[{2 \cdot \left( \Phi_0 + RM \cdot \frac{c}{\nu} \right)^2}\right]\\
V & = 0 ,
\end{align}
where $I$,$Q$,$U$, and $V$ are the flux densities of the Stokes parameters, $c$ is the speed of light, and $\nu$ the frequency of the channel, and using the relations for the linear feeds of the WSRT
\begin{equation}
XX = I + Q \quad XY = U + V \quad YX = U - V \quad YY = I - Q
\end{equation}
the fluxes for all four cross-correlations could easily be calculated for each individual channel. A summary of the parameters used is listed in Tab. \ref{table_calflux}.

\begin{figure}
	\resizebox{\hsize}{!}{\includegraphics{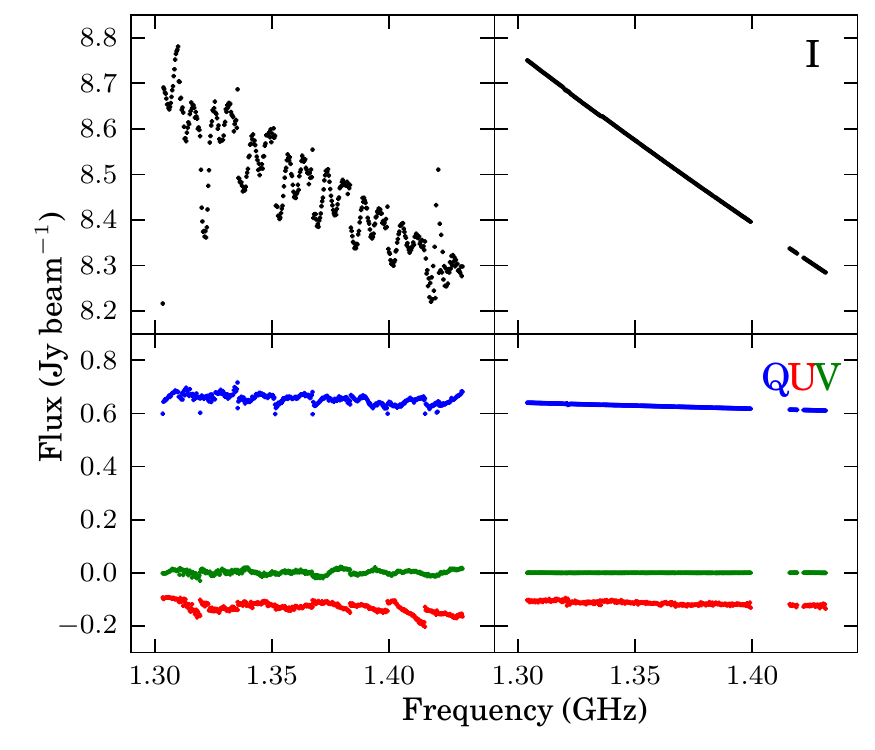}}
	\caption{Flux of the central pixel of the used calibrator 3C138. The top plots show the flux in Stokes I and the bottom ones the flux in Stokes Q, U, and V. The improvement between the subband based calibration (left) and the channel based one (right) is obvious.}
	\label{plot_polcal}
\end{figure}

This channel based derivation of the calibrator gains minimises the effects of the 17\,MHz-ripple in the pointing centre of the beam and reduces the instrumental polarisation significantly, which can vary by up to 1.5\% across the frequency band \citepads{2008A&A...489...69B}. Fig.\ref{plot_polcal} shows the difference between assuming a constant gain across one subband and the channel-based calibration performed here.

After this initial calibration step, each dataset was self-calibrated with different strategies described in the next subsections.

\subsubsection{18\,cm and 22\,cm datasets}

The first 22\,cm dataset contained a single 12\,hr run with a total bandwidth of 80\,MHz split into eight subbands centered at 1450, 1424, 1419, 1398, 1374, 1349, 1326, and 1303\,MHz. Each subband was covered by 128 channels in two cross-correlations. The integration time was 30 seconds, which was sufficient to sample the ionospheric fluctuations. The observed calibrators for this dataset are 3C147 and 3C295. The second dataset contained a complete 12\,hr run with frequency switching between 18\,cm and 22\,cm every 5 minutes yielding an integration time of $\sim$ 360 minutes for each frequency band. The subbands were centred at 1753, 1737, 1721, 1705, 1689, 1673, 1657, and 1641\,MHz for the 18\,cm-band and at 1422, 1406, 1390, 1374, 1358, 1342, 1326, and 1310\,MHz for the 22\,cm-band. Each subband was covered by 64 channels in all four cross-correlations with an integration time of one minute. The observed calibrators for these two datasets were 3C138 and CTD93.

After applying the calibrator gains with the strategy described in Sect. \ref{text_channel_based}, both datasets were exported to MIRIAD for faster imaging and cleaning. Due to the extremely luminous extended core of M82, a special self-calibration technique, described in Sect. \ref{text_cal_strong_source}, had to be used to reach a dynamic range of $\sim$50000. For each step in this data reduction involving cleaning masks were used from previous data reductions to include all visible flux inside the field.

Because of the better data quality of the observations with 80\,MHz bandwidth, the shorter integration time and the similar subband frequencies, clean models from the former dataset at appropriate frequencies were used to start the self-calibration process for the dataset with 160\,MHz bandwidth at 22\,cm. This lead to 13 usable images, which were then combined with an inverse square weighting and primary beam corrected.

The 18\,cm dataset was only used for polarisation purposes and therefore no total power images were produced.

\subsubsection{Calibration technique for strong extended sources}
\label{text_cal_strong_source}

The East-West alignment of the WSRT and the observation mode with only one calibrator observation in the beginning and one at the end makes self-calibration the most important step in the data reduction for high dynamic-range images. One has to make sure that especially the first models during the self-calibration process do not include spurious features, which would then be transported into images and models for later stages. These features are normally easily visible in the images since they are biased towards calibrator observations and array layout. This means for the strong extended source M82 that artificial clean components would be used which elongate the source in the East-West direction.

\begin{figure*}
	\resizebox{\hsize}{!}{\includegraphics{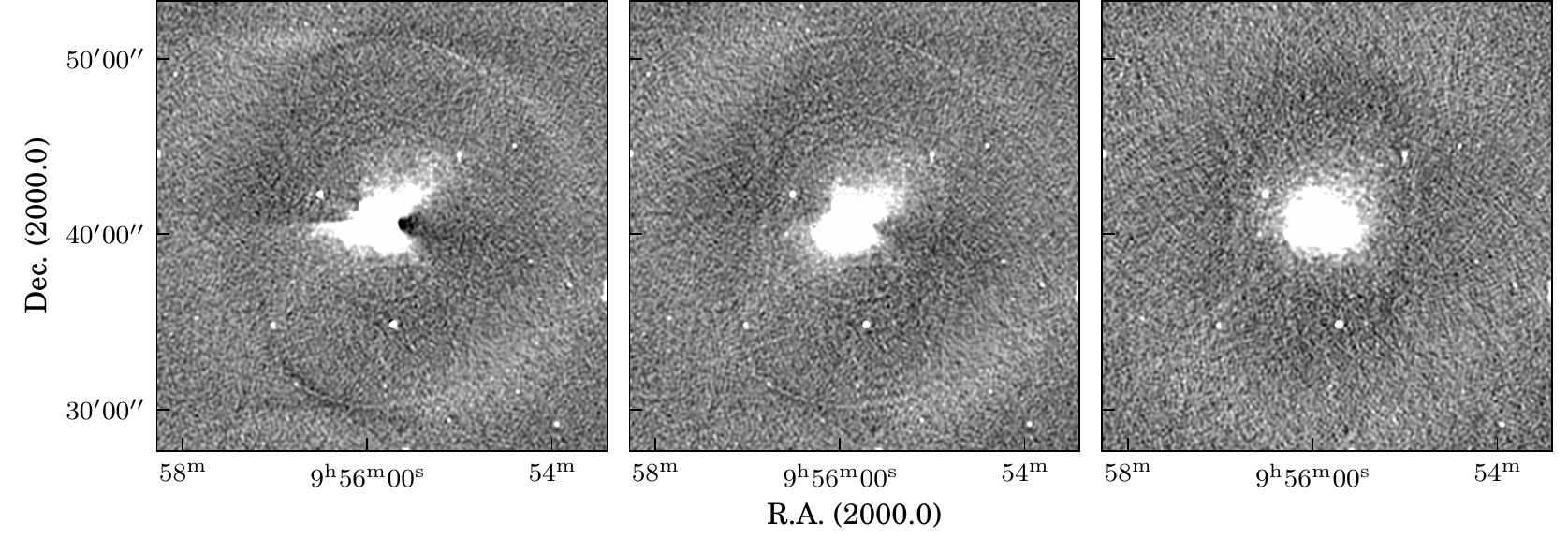}}
	\caption{A comparison of the image quality using a subband-based initial calibration technique and standard self-calibration procedure (left), a subband-based initial calibration and the self-calibration technique described in Sect. \ref{text_cal_strong_source} (middle), and a channel-based initial calibration technique and the self-calibration technique described in Sect. \ref{text_cal_strong_source} (right). The biasing of the flux towards the East- and West-direction and resulting sidelobes are easily visible in the first image. These effects nearly vanish in the second image and completely disappear in the last one. The images show only the first subband of the observations with 80\,Mhz bandwidth at $\uplambda$22\,cm without any primary beam correction.}
	\label{image_speccal}
\end{figure*}

Unresolved sources are normally very good calibrators since their total emission is covered by a clean beam. This makes them much better sources for self-calibration than strong extended sources since their clean component model is only one delta-function and not multiple delta-functions resembling the complex diffuse emission like it is for M82-type sources. The clean components for these sources would resemble the artificial elongation in the East-West-direction. Therefore, the best choice would be to construct a (u,v)-dataset of the M82-field with a dominating flux contribution from point sources. This was achieved by using the following technique:

\begin{enumerate}
	\item The channel-based initial calibration (Sect. \ref{text_channel_based}) was applied to the dataset. This minimises the contribution of the 17\,MHz ripple in the central part of the field.
	\item We applied a normal self-calibration approach for the entire field using only baselines $\geq4$\,k$\uplambda$ for the first self-calibration loop with a high solution interval. This parameter was reduced as well as the solution interval during further loops.
	\item We subtracted M82 from the dataset using the gains and the clean model from the last self-calibration loop. Afterwards the inverted gains from the same loop were applied resulting in a dataset without M82 and only the gains from the initial calibration applied.
	\item A self-calibration on this dataset was applied using only the point sources yielding an average gain over the entire field. After two rounds of self-calibration with long solution intervals and (u,v)-ranges $\geq4$\,k$\uplambda$ the gains were transferred to the original initially calibrated dataset containing all point-sources and M82. Then the self-calibration was applied to this dataset in the normal manner reducing the solution interval and including smaller (u,v)-ranges to improve the model.
\end{enumerate}

This technique enabled us to use the high-signal-to-noise-ratio of the strong extended source without artificially including clean components from the East-West-biasing. A comparison of the image quality before and after the described technique was applied is shown in Fig. \ref{image_speccal}.

\subsubsection{Polarisation calibration}

Due to high rotation measures already detected by \citetads{1994A&A...282..724R} the derivation of a model for all four Stokes parameters for each single subband is not sufficient. The Stokes Q and U fluxes vary significantly over the bandwidth of one subband. This would result in a wrong model for a significant number of channels in each subband, shifting flux from Q to U or vice versa and producing artificially false rotation measure components. Therefore, self-calibration for each single channel was done independently as described by \citetads{2005A&A...441..931D}. The high flux of M82 was sufficient to reach the signal-to-noise required for a satisfactory dynamic range in the Stokes Q- and U-parameters.

The whole calibration was done in MIRIAD using a script and including all (u,v)-ranges in the self-calibration. Masks from the total power data reduction were used to include all diffuse emission in the model. The final self-calibration loop used an amplitude and phase calibration with a solution interval of one minute. The final images were then cleaned to two times the noise level and primary beam corrected. After that all images were inspected for significant artefacts due to calibration errors and leftover RFI by eye. At the time of the observation the sun was at a maximum of its activity, which led to significant signals in the polarisation images far off the pointing centre. Since these signals are heavily time- and frequency-dependent, a subtraction of the signal was not possible, so that these images had to be discarded, but images at frequencies with a lower contribution of the signal from the sun were usable. Additionally, any images suffering from a significant elongation of the beam as a result of time-based flagging of the data of an east-west-interferometer like the WSRT were not used. Images with a noise floor three times higher than the average were also excluded. This yielded only 396 (187 for the $\uplambda$18\,cm and 209 for the $\uplambda$22\,cm observations) out of 1024 final images for each polarisation product.

\subsubsection{92\,cm dataset}

The 92\,cm dataset contained a single 12\,hr exposure with a total bandwidth of 20\,MHz. Each of the 8 subbands was sampled by 128 channels in two cross correlations. The central subband frequencies were set to 321, 325, 332, 336, 372, 376, 380, and 385 MHz. The observed calibrators were 3C147 and 3C295. The integration time was 30 seconds, which is sufficient to sample the ionospheric fluctuations during the night. Unfortunately some part of the observation was carried out during the day and at sunset, where the interference from the sun is especially high on the short baselines. Subtracting a clean model of the sun after the initial calibration was not successful, so that several hours on the shorter baselines had to be flagged.

The data was then exported to MIRIAD and self-calibration was carried out for each subband independently using masks during the clean step to include all diffuse emission in the field. The small bandwidth of each single subband of only 2.5 MHz made it possible to use a normal Clark-clean algorithm rather than a multi-frequency deconvolution \citepads{1994A&AS..108..585S} to reach a dynamic range of $\sim$10000. The subband images were then combined with an inverse square weighting and primary beam corrected.

\section{Total power emission}
\label{text_totalpower}

\subsection{Morphology}

\begin{figure}
	\resizebox{\hsize}{!}{\includegraphics{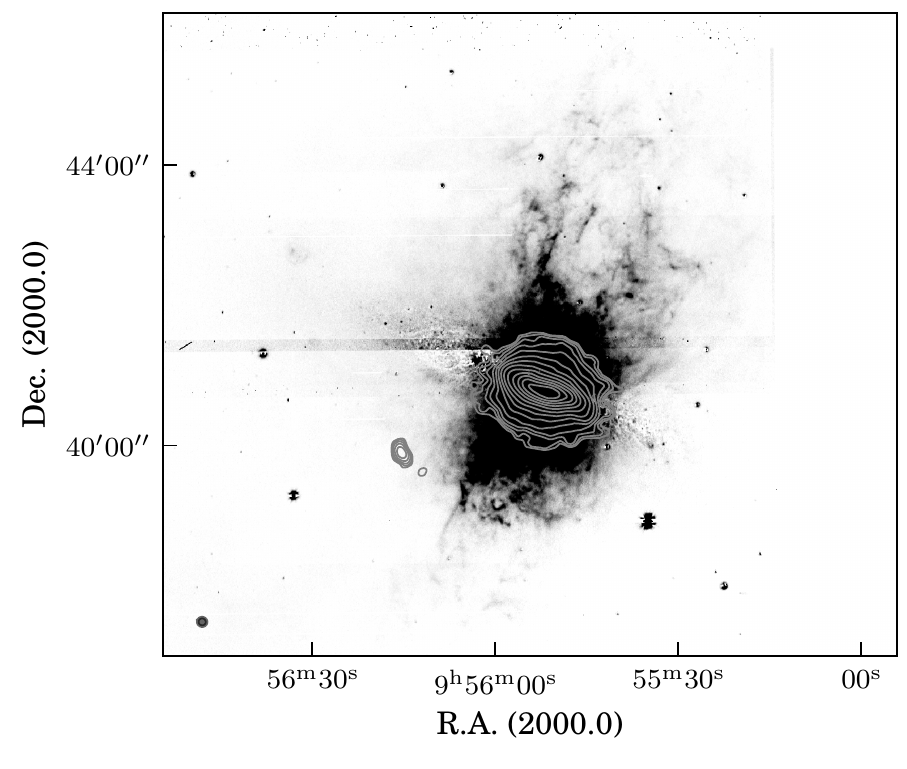}}
	\caption{Total power radio continuum contours at $\uplambda$3\,cm from the VLA-observations overlaid on a H$\upalpha$ image from the WIYN 3.5\,m Observatory provided by M. Westmoquette et al. (priv. comm.). Contours start at a 3$\sigma$ level of 50\,$\upmu$Jy/beam and increase in powers of 2. The beam size is 7\farcs6$\times$7\farcs3 and is shown in the bottom left corner of the image.}
	\label{image_TP_3cm}
\end{figure}

\begin{figure}
	\resizebox{\hsize}{!}{\includegraphics{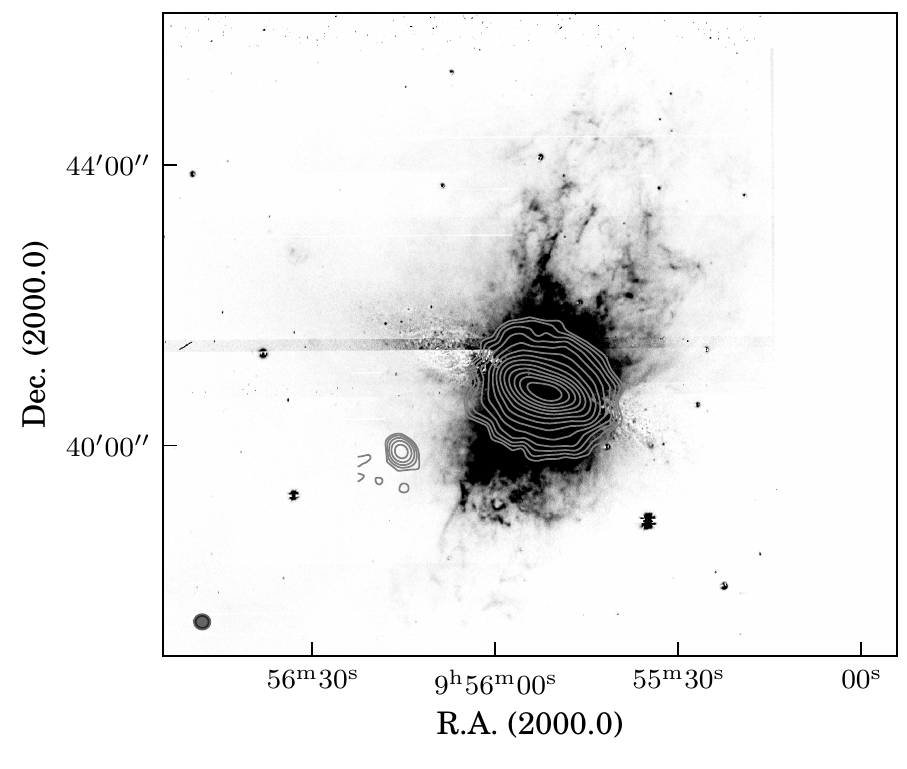}}
	\caption{Total power radio continuum contours at $\uplambda$6\,cm from the VLA-observations overlaid on a H$\upalpha$ image from the WIYN 3.5\,m Observatory provided by M. Westmoquette et al. (priv. comm.). Contours start at a 3$\sigma$ level of 120\,$\upmu$Jy/beam and increase in powers of 2. The beam size is 12\farcs5$\times$11\farcs7 and is shown in the bottom left corner of the image.}
	\label{image_TP_6cm}
\end{figure}

\begin{figure}
	\resizebox{\hsize}{!}{\includegraphics{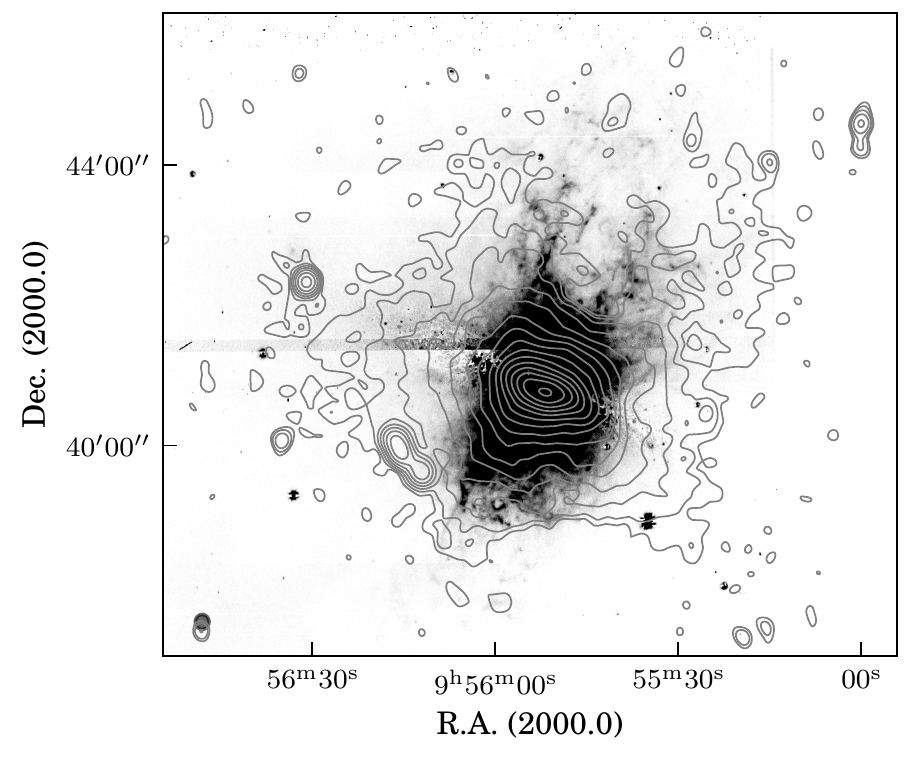}}
	\caption{Total power radio continuum contours at $\uplambda$22\,cm from the WSRT-observations overlaid on a H$\upalpha$ image from the WIYN 3.5\,m Observatory provided by M. Westmoquette et al. (priv. comm.). Contours start at a 3$\sigma$ level of 90\,$\upmu$Jy/beam and increase in powers of 2. The beam size is 12\farcs7$\times$11\farcs8 and is shown in the bottom left corner of the image.}
	\label{image_TP_22cm}
\end{figure}

\begin{figure}
	\resizebox{\hsize}{!}{\includegraphics{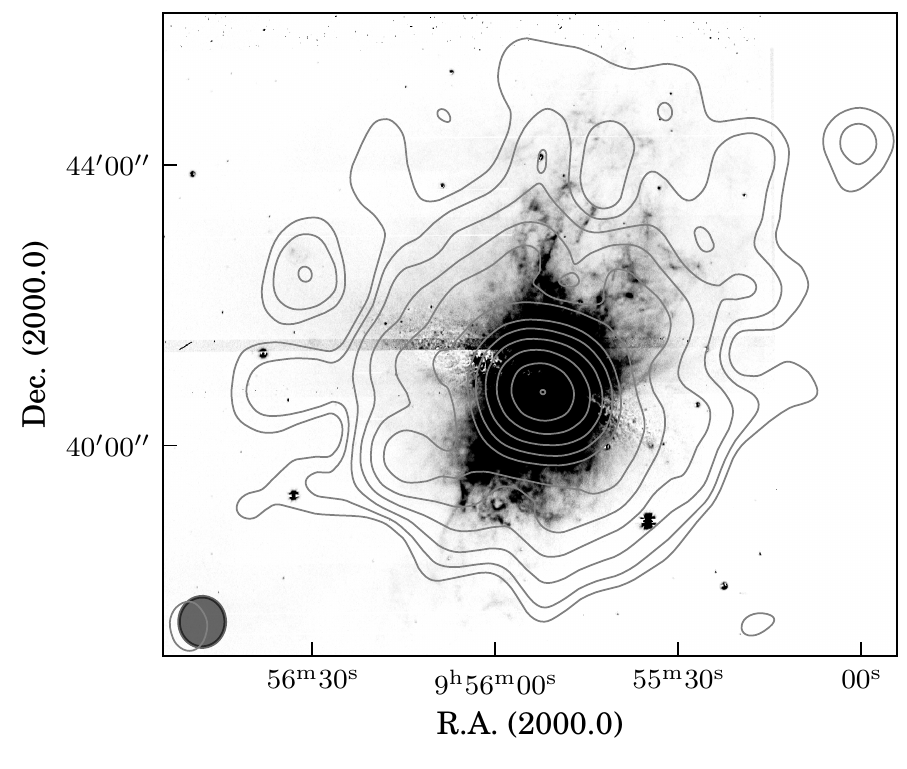}}
	\caption{Total power radio continuum contours at $\uplambda$92\,cm from the WSRT-observations overlaid on a H$\upalpha$ image from the WIYN 3.5\,m Observatory provided by M. Westmoquette et al. (priv. comm.). Contours start at a 3$\sigma$ level of 1.8\,mJy/beam and increase in powers of 2. The beam size is 43\farcs1$\times$39\farcs6 and is shown in the bottom left corner of the image.}
	\label{image_TP_92cm}
\end{figure}

The maps showing the total power emission are presented in Fig. \ref{image_TP_3cm}, Fig. \ref{image_TP_6cm}, Fig. \ref{image_TP_22cm}, and Fig. \ref{image_TP_92cm} at wavelengths of 3\,cm, 6\,cm, 22\,cm, and 92\,cm, respectively.

The morphology of the total power emission depends strongly on the observed wavelength. While the maps at 3\,cm and 6\,cm show only emission in the central disk and some extraplanar emission to the north and south of the very central region of M82, the maps at 22\,cm and 92\,cm show the structure of the radio halo and the disk emission. Overall the halo emission in the southern part seems to be less extended than in the northern part, but the whole halo emission correlates with the H$\upalpha$- as well as the PAH$^+$-morphology \citepads{2010A&A...514A..14K}.

The most obvious difference between the northern and southern parts of the halo are two extensions accompanied by a suppression of emission in between them in the northern halo. While the 92\,cm map shows basically the same features towards the north and south with a smaller extent towards the south, the 22\,cm morphology is completely different in the southern halo. Here the morphology is dominated by an extension south of the starbursting core with lacking emission in the southwestern part.

While the extent of the southern extensions at $\uplambda$92\,cm is about the same, it is different in the northern halo at $\uplambda$22\,cm as well as at $\uplambda$92\,cm. Here the northwestern spur seems to be more extended than the northeastern. \citetads{2002MNRAS.334..912M} already mentioned a difference in the distribution of compact radio sources in the core region between the eastern and western part. The western part seems to be denser populated by supernova remnants than by \ion{H}{ii} regions while it is vice versa in the eastern part. This could mean that the western part is further evolved and has ejected the material into the halo earlier than the eastern part where a higher supernova activity is expected within the next Myrs.

Remarkable is the morphology of the radio emission along the disk of M82. Observations by \citetads{1991ApJ...369..320S} and \citetads{1992A&A...256...10R} already showed this extent of the emission. Therefore it is most likely that the radio morphology of M82 consists of a superposition of a disk and a biconical outflow.

\begin{table}
	\caption{Integrated fluxes for our M82 observations}
	\centering
	\begin{tabular}{@{} cc @{}}
		\toprule
		\toprule
		$\uplambda$ (cm) & Flux (Jy) \\
		\midrule
		3 & $2.20\pm0.11$ \\
		6 & $3.23\pm0.16$ \\
		22 & $7.41\pm0.37$ \\
		92 & $13.83\pm0.69$ \\
		\bottomrule
	\end{tabular}
\end{table}

While the lowest contour levels for the 3\,cm, 6\,cm, and 22\,cm maps have similar noise levels, the large halo emission is only visible in the 22\,cm and 92\,cm map. One could argue that this might be partly due to lower sensitivities of the $\uplambda$3\,cm and $\uplambda$6\,cm maps to extended emission because of missing spacings. Comparing our integrated flux of 3.23\,Jy\,$\pm$\,0.16\,Jy with the single dish observations of \citetads{1991ApJS...75.1011G} at $\uplambda$6\,cm, who measured a flux of 3.96\,Jy\,$\pm$\,0.60\,Jy, may support this statement. The surrounding point-sources were not subtracted for the single-dish measurement due to the lacking resolution, but adding up the flux of all point-sources in our interferometer image within the beam of the single dish telescope yields only $\sim20\,$mJy. 

Hence, even if this is less than the observed difference of the two measurements, it brings the values in agreement within their errors. The largest observable scale of the VLA at $\uplambda$3\,cm is $3\arcmin$ and $4\arcmin$ at $\uplambda$6\,cm in comparison to the WSRT with $21\arcmin$ at $\uplambda$22\,cm and 81$\arcmin$ at $\uplambda$92\,cm. This is similar in size to the observed structures at $\uplambda$3\,cm and $\uplambda$6\,cm respectively and larger than the observed structures at $\uplambda$22\,cm and $\uplambda$92\,cm. However, the rapid decrease of our emission at $\uplambda$3\,cm and $\uplambda$6\,cm in direction away from the galaxy is stronger than expected from missing spacings alone.

\citetads{2011A&A...531A.127S} also used the VLA D-array configuration for their studies of NGC5775 and could recover emission with a size of 3\farcm5 at $\uplambda$3\,cm and $\uplambda$6\,cm. This is two times larger in comparison to our structure size of $1.5\arcmin\sim2\arcmin$. A combination of their VLA measurements with single-dish data from the Effelsberg telescope recovered only 3\,\% more flux at $\uplambda$3\,cm and no increase was recognised at $\uplambda$6\,cm.  Therefore we conclude that missing spacings do not significantly influence the integrated fluxes or the overall morphology.

\subsection{Total intensity distribution}
\label{total_intensity_distribution}

To isolate the emission of the galaxy from contributions of background sources, all point-like sources exceeding three times the rms noise in the images were fitted with Gaussians in the image plane and subtracted. This method was successful for all sources except the double source to the southeast of M82 due to its complicated structure and the fact that it is embedded in the diffuse emission. Therefore this source was left in the images and for parts of the data analysis where its flux contributes, the data points were not included in the analysis.

\subsubsection{Radial total intensity profile}

\begin{figure}
	\resizebox{\hsize}{!}{\includegraphics{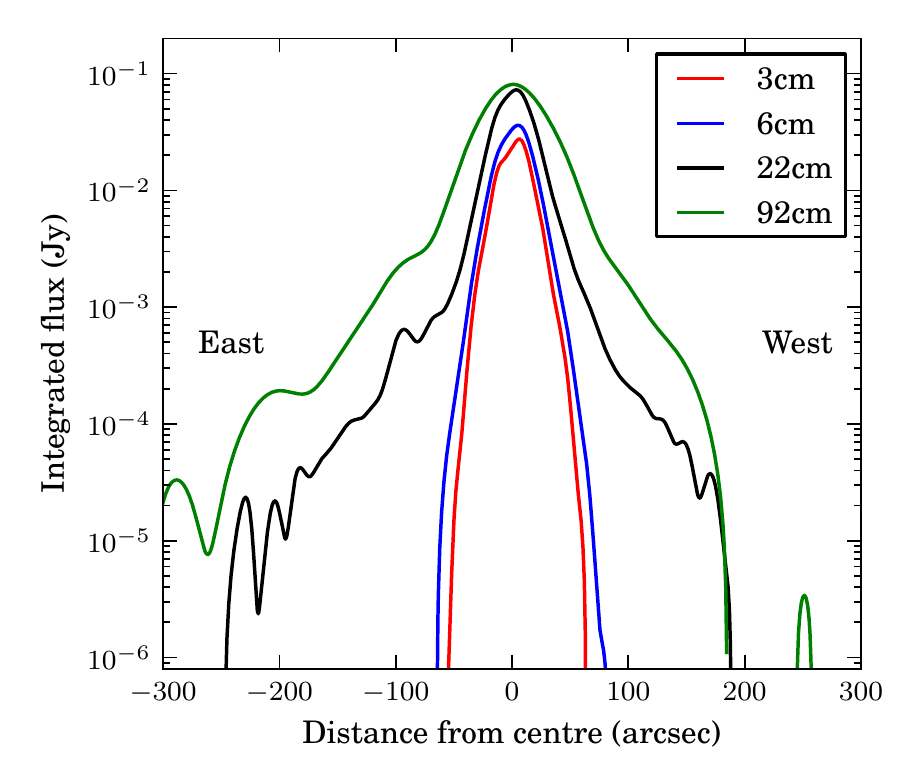}}
	\caption{Integrated total intensity for each individual pixel column plotted versus the distance from the centre of the galaxy along the major axis. The flux was integrated over the total power emission along the minor axis of the galaxy. The maximum at -100\arcsec~is the consequence of the source left inside the diffuse emission (see Sect. \ref{total_intensity_distribution}).}
	\label{plot_TP}
\end{figure}

The resulting total intensity profiles along the major axis are shown in Fig. \ref{plot_TP} for all four observed wavelengths. The individual values were obtained by regridding the total power maps to a SLOAN g' image of the galaxy and integrating the flux over each pixel column. Since we are only interested in the overall shape of the profile and the relative differences between the wavelengths, the absolute integrated fluxes are irrelevant.

While the emission in the central part is dominated by the star-forming nucleus of M82, the 22\,cm- and 92\,cm-observations show significant emission in the disk of the galaxy. Remarkable is a slight shift of the emission maximum to the west visible in the higher frequency images due to their better resolution. This seems to be a consequence of the higher star-forming activity in the western part of the central region of the galaxy \citepads{1994MNRAS.266..455M} and might even cause the difference between the extent of the spurs in the northern region. The most extended feature reaching up to $\sim$4.5\,kpc into the halo is positioned directly north of this region. Here the cosmic ray gas may be able to escape the galaxy easier since the increased star-formation simply blew away the dense medium providing a channel towards which the gas can escape from the central starburst.

\subsection{Vertical scaleheights}

The observed radio emission of an inclined galaxy is a mixture of the halo emission and the projected disk emission contributing to the halo emission \citepads{1995A&A...302..691D}. In order to simulate the latter, we compressed the intensity distribution along the major axis by the inclination of the galaxy. The resulting distribution was then convolved with the HPBW of the observation to account for beam smearing effects. The FWHM of a Gaussian fitted to the major axis profile of this image is then the effective beam size. This led to values of 16\arcsec (265 pc) at $\uplambda$3\,cm, 19\arcsec (330 pc) at $\uplambda$6\,cm, 21\arcsec (350 pc) at $\uplambda$22\,cm, and 50\arcsec (860 pc) at $\uplambda$92\,cm. All emission outside this distance from the major axis can be attributed to extraplanar emission.

To analyse the propagation of cosmic rays and their losses a parameter, which is independent of the total flux, is needed. Therefore the vertical scaleheights were used. This approach has already been performed for other objects \citepads{1998LNP...506..555D,2009A&A...494..563H,2011A&A...531A.127S} and makes the values comparable to each other.

According to \citetads{1995A&A...302..691D} the fitted function in case of a Gaussian distribution is
\begin{equation}
W_{gauss}(z) = \frac{w_0 z_0}{\sqrt{2\sigma^2 + z_0^2}}\exp\left({-\frac{z^2}{2\sigma^2+z_0^2}}\right)
\end{equation}
with $\sigma$ being the effective beam size, $w_0$ the maximum of the distribution and $z_0$ the scaleheight. In case of an exponential distribution this becomes
\begin{equation}
\begin{split}
W_{exp}(z) =\, & \frac{w_0}{2}\exp{\left(-\frac{z^2}{2\sigma^2}\right)} \\
& \cdot \left[ \exp{ \left (\frac{\sigma^2 - zz_0}{\sqrt{2}\sigma z_0} \right)}^2 \text{erfc} \left( \frac{\sigma^2 - zz_0}{\sqrt{2}\sigma z_0} \right) \right. \\
& \left. + \exp{ \left( \frac{\sigma^2 + zz_0}{\sqrt{2}\sigma z_0} \right)^2 } \text{erfc} \left( \frac{\sigma^2 + zz_0}{\sqrt{2}\sigma z_0} \right) \right]
\end{split}
\end{equation}
with the complementary error function:
\begin{equation}
\text{erfc}\,x = 1 - \text{erf}\,x = \frac{2}{\sqrt{\pi}} \int\limits_x^{\infty} \exp(-r^2)\,dr
\end{equation}
Integrations over rectangular areas with a width of 60\arcsec parallel to the galaxy's minor axis were used as an input to a Gaussian, a one component exponential and a two component exponential fit. This divides the galaxy into 3 strips each north and south of the major axis for the $\uplambda$22\,cm and $\uplambda$92\,cm data and one strip each for the $\uplambda$3\,cm and $\uplambda$6\,cm data resembling a good database for the spurs in the east and west and the outflow cone in the centre. This integration technique ensures that the fits do not suffer from small scale fluctuations in the images.

We used the fitting routines by \citetads{1995A&A...302..691D}, which take the calculated effective beam size into account. Our measured values and the three different types of fit functions are presented in Fig. \ref{plot_sh_3cm} to Fig. \ref{plot_sh_92cm}, the deprojected scaleheights are presented in Tab. \ref{table_sc}.

Data points where the background radio source in the southwest contributed to the diffuse emission were not used for the fitting routines. Errors $\sigma_S$ for the data points were calculated using:
\begin{equation}
\sigma_S = \sqrt{\sigma_{rms}^2 + \sigma_B^2},
\end{equation}
where $\sigma_{rms}$ is the error due to the noise in the image, which was estimated by selecting an emission free area close to the centre of the image and calculating the standard deviation, and $\sigma_B = \sigma_{rms}\sqrt{n}$ the calculated baselevel error in the image, with $n$ being the number of integrated beam areas. Additionally the standard deviation inside each integration area was calculated and in cases where it was higher than the calculated baselevel error, $\sigma_B$ was replaced by this value. This is especially critical in areas with strong gradients like the core region, where the actual fluctuations of the source over the area is dominating the error for the integration.

\subsubsection{Scaleheights at $\uplambda$3cm and $\uplambda$6cm}

\begin{figure}
	\resizebox{\hsize}{!}{\includegraphics{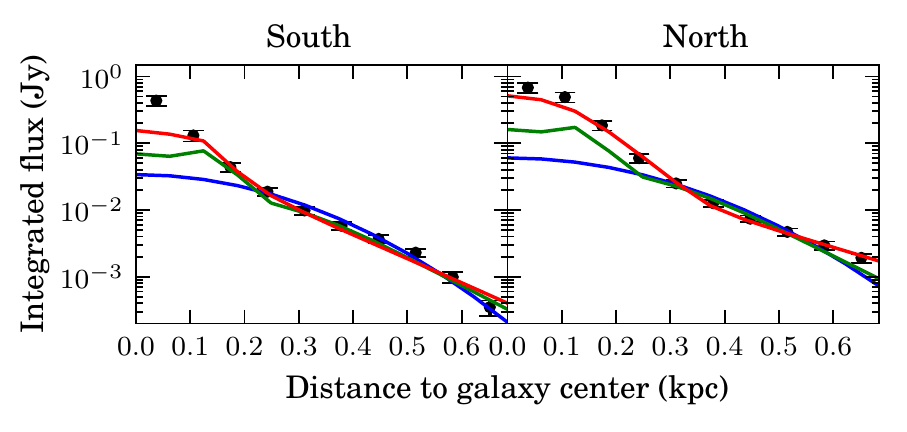}}
	\caption{Intensity distribution and its errors for the $\uplambda$3\,cm-data. Each data point reflects an integration over an area of 60''$\times$4.5''. The blue line shows a Gaussian fit, the green line a one component exponential fit, and the red line a two component exponential fit. The parameters for the fits are listed in Tab. \ref{table_sc}.}
	\label{plot_sh_3cm}
\end{figure}

\begin{figure}
	\resizebox{\hsize}{!}{\includegraphics{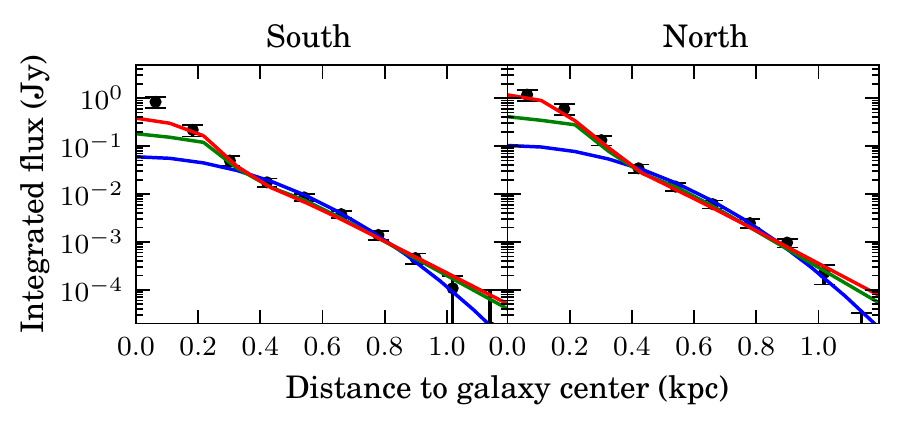}}
	\caption{Intensity distribution and its errors for the $\uplambda$6\,cm-data. Each data point reflects an integration over an area of 60''$\times$7''. The blue line shows a Gaussian fit, the green line a one component exponential fit, and the red line a two component exponential fit. The parameters for the fits are listed in Tab. \ref{table_sc}.}
	\label{plot_sh_6cm}
\end{figure}

Since the observations at $\uplambda$3\,cm and $\uplambda$6\,cm only show emission in the central part of the galaxy, the analysis of the scaleheights is mostly limited to the disk and inner halo emission. Fig. \ref{plot_sh_3cm} and Fig. \ref{plot_sh_6cm} clearly show that the one component and two component exponential fits fit much better than the Gaussian fit.

The absolute values for the $\uplambda$3\,cm data and $\uplambda$6\,cm data do not differ a lot from each other. This is a clear sign for a homogeneous halo property at these wavelengths for this part of the galaxy. Therefore an average value of the scaleheights can be used by simply calculating the mean. This yields average scaleheights of $20\pm5$\,pc for the thin disk and $140\pm30$\,pc for the thick disk at $\uplambda$3\,cm and $\uplambda$6\,cm.

The influence of missing spacings to these values is assumed to be negligible since the fits are weighted by the relative error of the individual data points. This error is normally scaling with the signal-to-noise ratio, which is much higher in the central part than it is for the diffuse outer regions, where missing spacings might influence the integrated flux values.

\subsubsection{Scaleheights at $\uplambda$22\,cm and $\uplambda$92\,cm}
\label{text_scaleheights_22_92}

\begin{figure}
	\resizebox{\hsize}{!}{\includegraphics{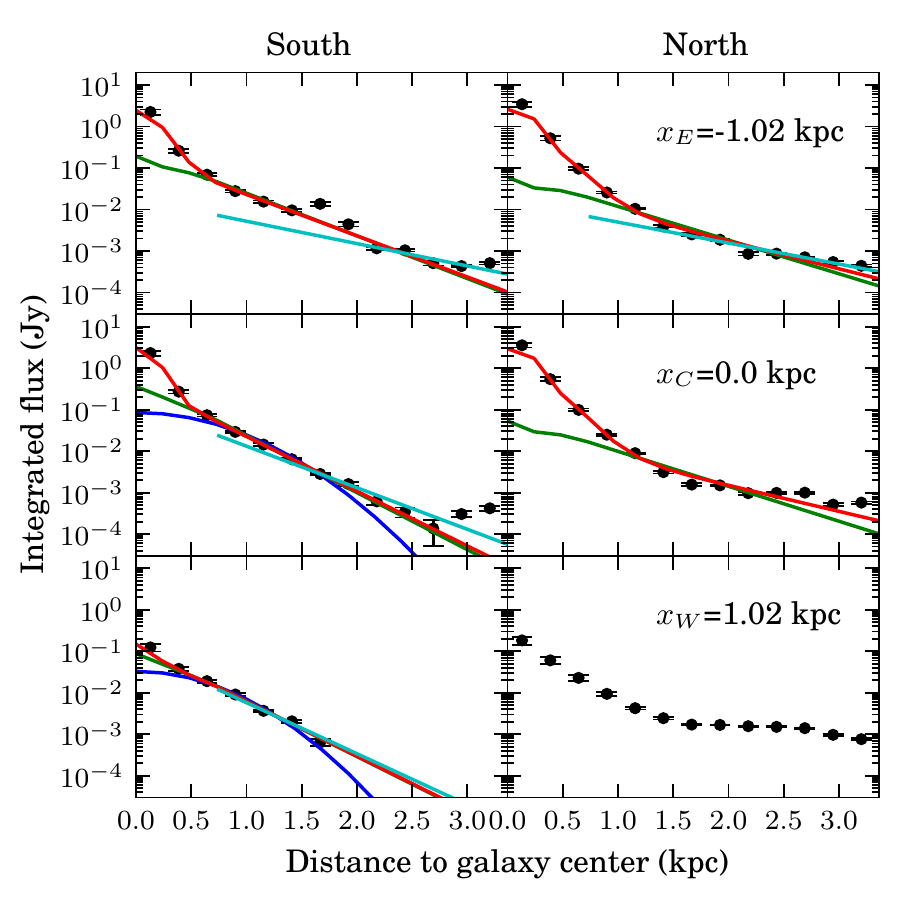}}
	\caption{Intensity distribution and its errors for the $\uplambda$22\,cm-data. Each data point reflects an integration over an area of 60''$\times$15''. The blue line shows a Gaussian fit, the green line a one component exponential fit, the red line a two component exponential fit, and the cyan line a one component exponential fit to the halo only. The parameters for the fits are listed in Tab. \ref{table_sc}.}
	\label{plot_sh_22cm}
\end{figure}

\begin{figure}
	\resizebox{\hsize}{!}{\includegraphics{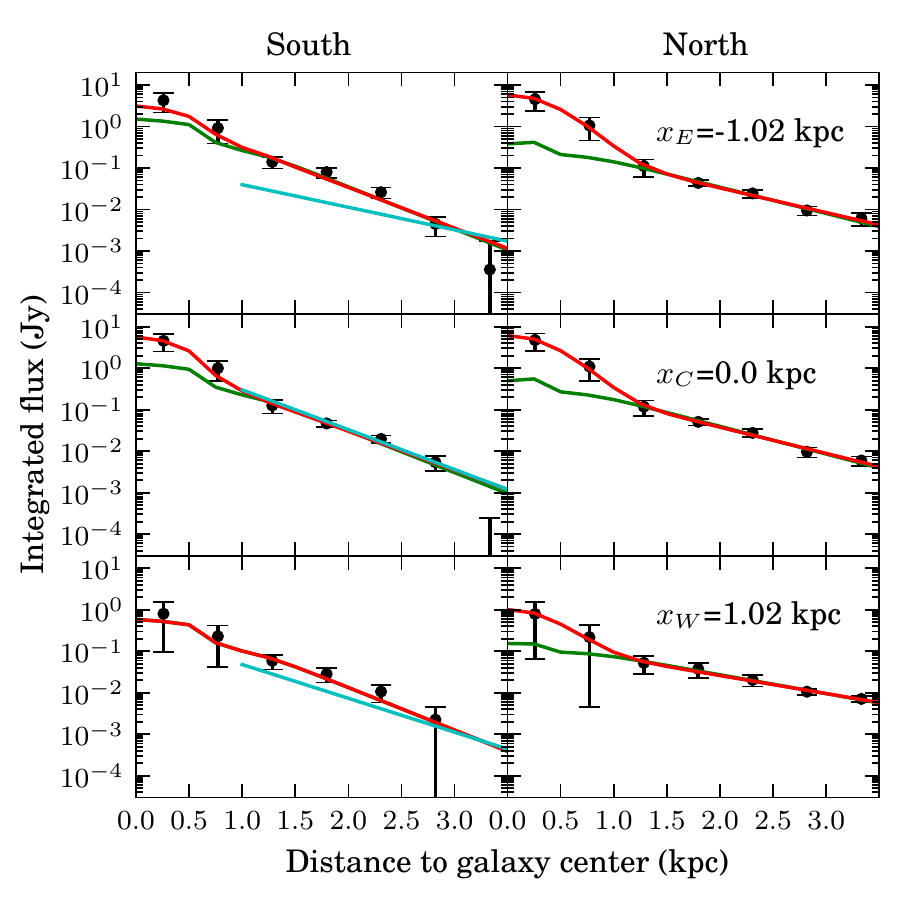}}
	\caption{Intensity distribution and its errors for the $\uplambda$92\,cm-data. Each data point reflects an integration over an area of 60''$\times$30''. The blue line shows a Gaussian fit, the green line a one component exponential fit, the red line a two component exponential fit, and the cyan line a one component exponential fit to the halo only. The parameters for the fits are listed in Tab. \ref{table_sc}.}
	\label{plot_sh_92cm}
\end{figure}

Looking at the emission profiles at longer wavelengths with high sensitivity shows more extended disk and halo emission. The fits in Fig. \ref{plot_sh_22cm} and Fig. \ref{plot_sh_92cm} clearly show that there is no function which fits all the data points, but for most cases the two component exponential fit is the best solution. The exception is the southwestern part of the galaxy, where the emission drops very rapidly in z-direction. This seems to be a clear sign of high losses of the cosmic ray electrons in this area. No successful fit could be found for the northwestern part at $\uplambda$22\,cm.

Remarkable is the difference in the scaleheights between the northern and southern part of the galaxy. Using again the mean of the $\uplambda$22\,cm values yields $690\,\pm\,50$\,pc for the northern thick disk and $390\,\pm\,70$\,pc for the southern one, and for the $\uplambda$92\,cm emission $820\,\pm\,170$\,pc and $450\,\pm\,20$\,pc respectively. Since most of the galaxies in the M81/M82-group core are located in the south of M82, the group medium is expected to be denser towards this direction. The reduced scaleheights in the southern part are a good argument for higher losses of cosmic rays towards the group centre. \citetads{2010A&A...514A..14K} detected a stronger radiation field only in the southeastern part of the galaxy which favours higher inverse Compton losses in this area, but this cannot explain the overall reduced extent of the southern halo and its reduced scaleheights. A slower southern outflow would be a possible explanation, but this would need a lower starburst activity and/or density in the southern central part to explain the reduced kinetic energy input. Arguing for a motion of M82 towards the group centre and a compression of the cosmic ray gas as has been seen in clusters of galaxies cannot be confirmed by simulations \citepads{1993egte.conf..253Y}, but is also not fully ruled out due to the uncertainties in the possible evolution scenarios.

Since it is most likely that secondary electrons are produced in the core of M82 and its surrounding region up to several hundred parsecs (see Sect. \ref{label_losses} for a detailed discussion) a one component exponential fit was also used to model loss processes that are independent of the core region. No strong difference is seen between the scaleheights at $\uplambda$22\,cm and $\uplambda$92\,cm. Averaging the scaleheights for the thick disk of both wavelengths results in $610\pm230$\,pc. All values for the thick disk in Tab. \ref{table_sc} are still by a factor of 2-4 smaller than the ones from normal spiral galaxies \citepads{1998LNP...506..555D,2011arXiv1111.7081K}.

\begin{table}
	\caption{Scaleheights between $\uplambda$3\,cm and $\uplambda$92\,cm for all successful fits. The indices in the first column show the position of the fit. $z_{0,2e,c}$ and $z_{1,2e,c}$ are the scaleheights for the first and second component to the two-component exponential fit and $z_{0,1e,h}$ is the scaleheight of the one-component exponential fit to the halo.}
	\label{table_sc}
	\centering
	\begin{tabular}{@{} ccccc @{}}
		\toprule
		\toprule
		Strip & $\uplambda$(cm) & $z_{0,2e,c}$(pc) & $z_{1,2e,c}$(pc) & $z_{0,1e,h}$(pc)\\
		\midrule
		$x_{S,C}$ & 3 & 16 & 121 \\
		$x_{N,C}$ & 3 & 22 & 181 \\
		\midrule
		$x_{S,C}$ & 6 & 22 & 134 \\
		$x_{N,C}$ & 6 & 20 & 129 \\
		\midrule
		$x_{S,E}$ & 22 & 72 & 445 & 819 \\
		$x_{N,E}$ & 22 & 146 & 663 & 874 \\
		$x_{S,C}$ & 22 & 40 & 335 & 441 \\
		$x_{N,C}$ & 22 & 143 & 705 \\
		$x_{S,W}$ & 22 & 20 & 334 & 359 \\
		$x_{N,W}$ & 22 & & & \\
		\midrule		
		$x_{S,E}$ & 92 & 114 & 450 & 811 \\
		$x_{N,E}$ & 92 & 107 & 745 & \\
		$x_{S,C}$ & 92 & 58 & 471 & 459 \\
		$x_{N,C}$ & 92 & 86 & 700 & \\
		$x_{S,W}$ & 92 & 431 & 431 & 541 \\
		$x_{N,W}$ & 92 & 77 & 1017 & \\
		\bottomrule
	\end{tabular}
\end{table}

\section{Spectral index distribution}
\label{text_spectralindex}

\subsection{Morphology}

\begin{figure}
	\resizebox{\hsize}{!}{\includegraphics{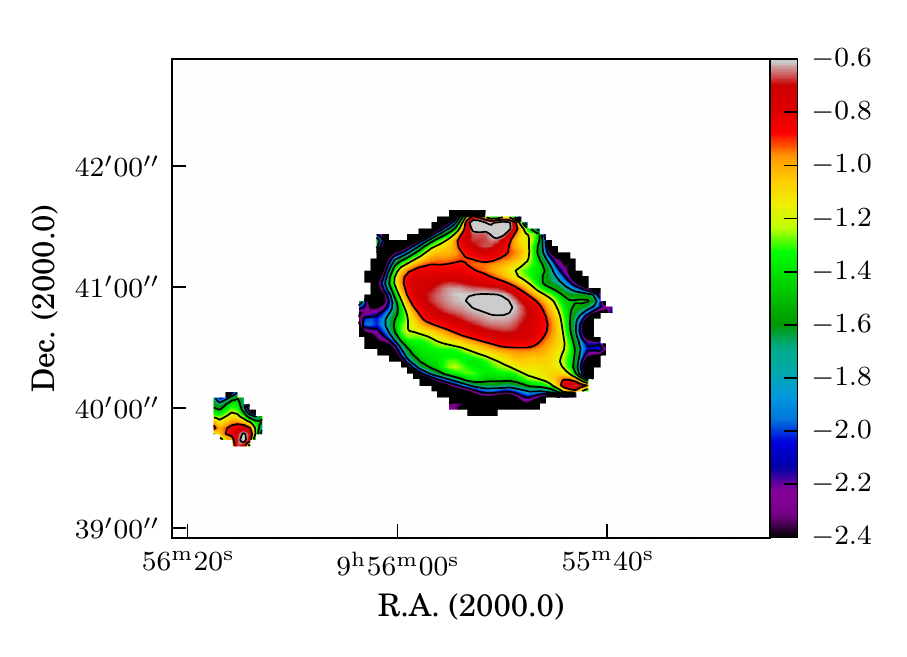}}
	\caption{Spectral index distribution between $\uplambda$3\,cm and $\uplambda$6\,cm. Contours start at -2.4 and end at -0.6 with an increment of 0.3.}
	\label{image_SI_3cm_6cm}
\end{figure}

\begin{figure}
	\resizebox{\hsize}{!}{\includegraphics{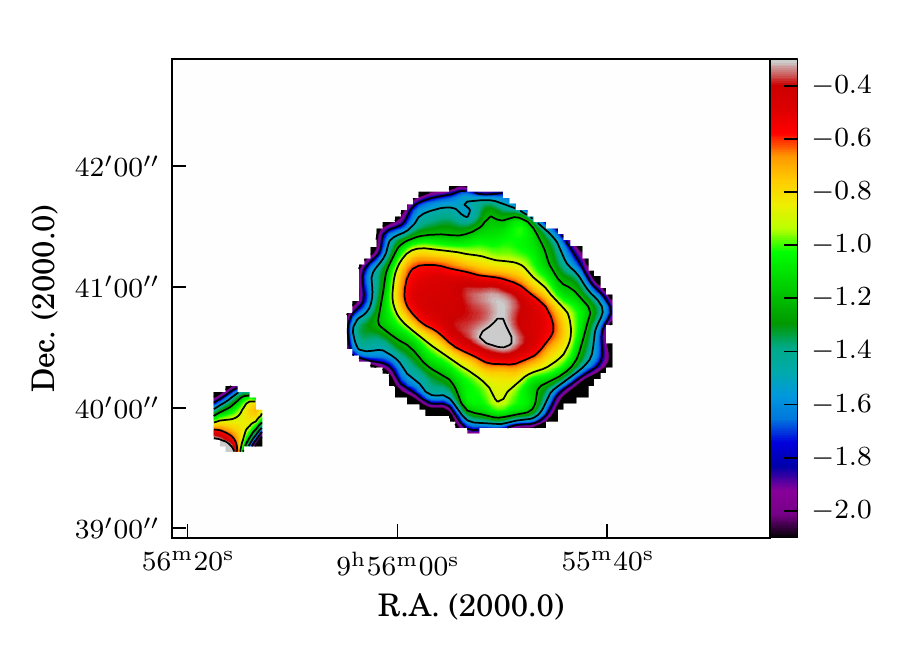}}
	\caption{Spectral index distribution between $\uplambda$6\,cm and $\uplambda$22\,cm. Contours start at -2.1 and end at -0.3 with an increment of 0.3.}
	\label{image_SI_6cm_22cm}
\end{figure}

\begin{figure}
	\resizebox{\hsize}{!}{\includegraphics{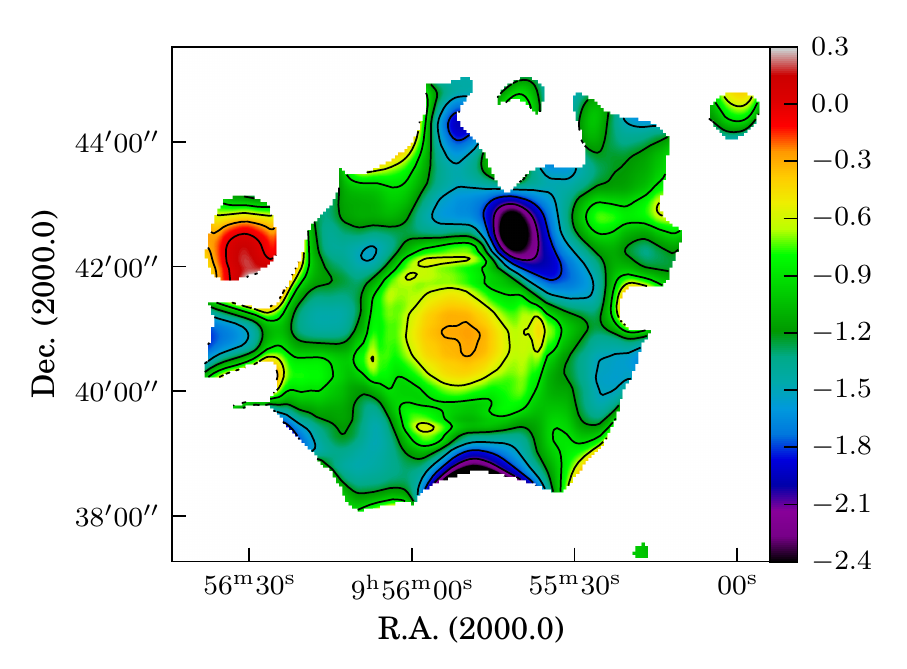}}
	\caption{Spectral index distribution between $\uplambda$22\,cm and $\uplambda$92\,cm. Contours start at -2.4 and end at +0.3 with an increment of 0.3.}
	\label{image_SI_22cm_92cm}
\end{figure}

Spectral index maps were produced between $\uplambda$3\,cm and $\uplambda$6\,cm (Fig. \ref{image_SI_3cm_6cm}), $\uplambda$6\,cm and $\uplambda$22\,cm (Fig. \ref{image_SI_6cm_22cm}), and $\uplambda$22\,cm and $\uplambda$92\,cm (Fig. \ref{image_SI_22cm_92cm}) using a 5$\upsigma$-cutoff level for the total power images. To check the contribution of different (u,v)-coverages for the different wavelengths, each spectral index map was once produced by limiting the used baselines to match the shortest and longest baselines of each other and once using all data and convolving the final maps with the same beam after image restoration and primary beam correction. No significant difference was visible so that the latter maps were used due to their higher sensitivity. Additionally, error images were calculated leading to errors of $\sigma_\alpha = 0.02$ for the core regions and up $\sigma_\alpha = 0.2$ for regions close to the cutoffs.

Thermal emission is known to contribute less than $10\%$ of the total flux density of M82 at $\uplambda$6\,cm and even does not dominate the spectrum at frequencies up to 25\,GHz \citepads{1991ApJ...369..320S,1988A&A...190...41K}, which implies a negligible thermal contribution to the measurements at $\uplambda22$ and $\uplambda92$\,cm.

All maps show an overall flat spectral index between $\alpha=-0.6$ and $\alpha=-0.8$ for the central part of the galaxy, which decreases towards steeper values of $\alpha=-1.5$ for the outer parts of the disk and halo emission. The outermost values in the maps between $\uplambda$3\,cm and $\uplambda$6\,cm and between $\uplambda$6\,cm and $\uplambda$22\,cm might be influenced by missing spacings. Therefore an overall steepening of the spectral index towards the halo cannot be neglected. The average spectral index gradient in the halo between $\uplambda$22\,cm and $\uplambda$92\,cm is relatively flat with an index of $\alpha=-1.2$ and does not steepen with increasing distance from the midplane (Fig. \ref{image_SI_22cm_92cm}). Two remarkable regions, one towards the northwest and one to the south, are visible between wavelengths of $\uplambda$3\,cm and $\uplambda$22\,cm, where the spectral index first steepens and then rises with increasing distance from the disk. Maps from \citetads{1991ApJ...369..320S}, who used different data for $\uplambda$22\,cm and $\uplambda$92\,cm, independently show the same characteristics at these locations.

Looking into these two regions at spectral indices derived from longer wavelengths does not show this behaviour due to the smaller beam size but following the same direction deeper into the halo shows spectral indices of $\alpha=-2.5$. While the northern region is located directly in the outflow cone indicating high inverse Compton, adiabatic, and/or synchrotron losses in this region, the southern region can be associated with higher inverse Compton losses towards a recently formed OB-association due to the stronger radiation field.

\subsection{Disentangling the core and halo emission}

\begin{figure}
	\resizebox{\hsize}{!}{\includegraphics{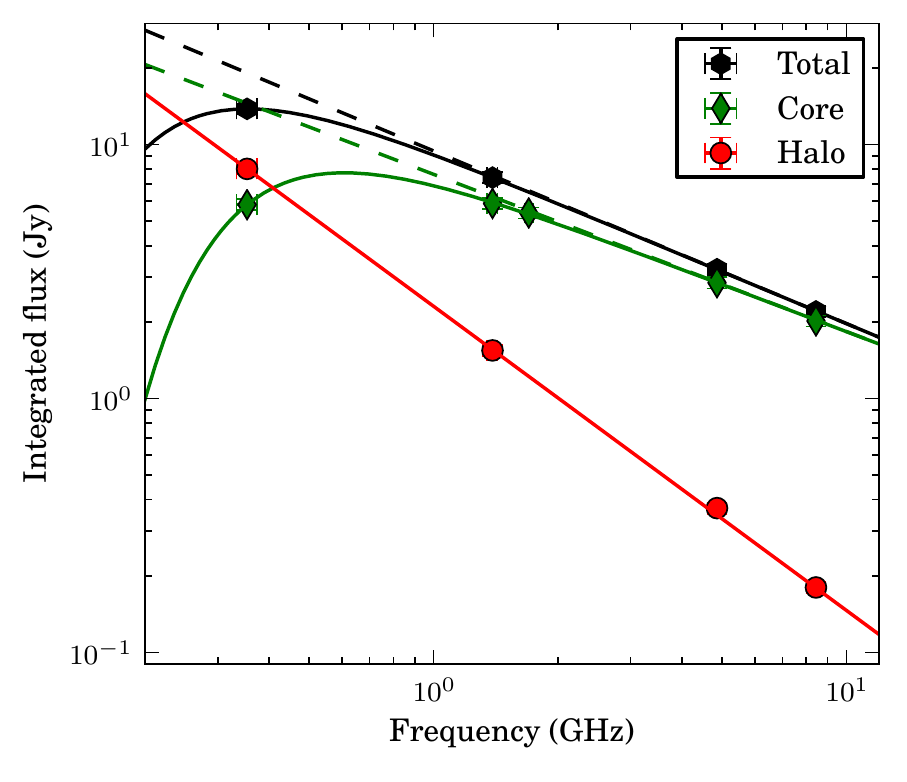}}
	\caption{Integrated flux of M82 for the whole galaxy, the core region and the halo plotted vs. the frequency. The additional point at $\uplambda$18\,cm is from \citetads{2007A&A...461..455B}. The integrated emission over the whole galaxy shows a turnover in the spectral index, which can be attributed to free-free absorption processes in the core region. The dashed lines represent the expected flux in case of no free-free absorption.}
	\label{image_SI}
\end{figure}

The loss processes of the cosmic ray electrons are expected to be different in the high density region of the core and in the halo. To analyse this the flux at each wavelength was integrated once over the whole galaxy, once in the starburst region with a typical radius of 450\,pc \citepads{2009AJ....137..517L} and once over the halo (Fig. \ref{image_SI}). An additional measurement could be acquired from \citetads{2007A&A...461..455B} for the core region, but the dynamic range of their maps was not sufficient to extract a measurement for the halo.

The spectral index $\alpha$ is defined as
\begin{equation}
S = S_0 \left(\frac{\nu}{\nu_0}\right)^\alpha.
\label{Eq:SI}
\end{equation}
The slope of the fits to the intensities $S$ in Jy at the wavelengths $\nu$ in GHz resembles the spectral index. Looking at the data points in a log-log-diagram (Fig. \ref{image_SI}) shows that only the halo component can be fitted with this function yielding $\alpha_h=-1.20\pm0.01$. Using only this function for the total emission resembles the average of the spectral index over the whole galaxy with $\alpha_a=-0.52\pm0.04$, but this is strongly influenced by the core region, where Eq. \ref{Eq:SI} does not provide an adequate fit.

A large amount of flux in the core region of M82 has been resolved into $\sim100$ independent discrete sources, which are a mixture of supernova remnants and ionised regions around newly forming star-clusters \citepads{2010MNRAS.408..607F,2002MNRAS.334..912M,1997MNRAS.291..517W}. The prior authors found indications that particles in these dense regions are suffering from free-free absorption processes inside these ionised regions and in the sources itself, which reduces the observed flux for low frequencies. Under the assumption that the measured flux for the resolution of the presented observations is dominated by these dense objects in the core region, the opacity $\tau$ of the medium can be included in Eq. \ref{Eq:SI} resulting in
\begin{equation}
S = S_0 \left(\frac{\nu}{\nu_0}\right)^\alpha e^{-\tau}
\label{Eq:SI_ff}
\end{equation}
in case of free-free absorption of sources inside an ionised medium and
\begin{equation}
S = S_0 \left(\frac{\nu}{\nu_0}\right)^2(1-e^{-\tau})
\label{Eq:SI_brems}
\end{equation}
in case of free-free absorption in the sources itself with
\begin{equation}
\tau = \frac{8.2\cdot 10^{-2} \nu^{-2.1} EM}{T_e^{1.35}},
\end{equation}
where $EM$ is the Emission Measure of the region in pc cm$^{-6}$ and $T_e$ the electron temperature of the warm medium in K \citepads{1997MNRAS.291..517W}. Assuming $T_e=10^4\,\text{K}$, the Emission Measure as well as the spectral index can be calculated by fitting Eq. \ref{Eq:SI_ff} and \ref{Eq:SI_brems} to the data. Only the fit for Eq. \ref{Eq:SI_ff} converged sufficiently, which suggests that sources surrounded by dense ionised star-forming regions are the origin for this low-frequency turnover. This results in a spectral index of $\alpha_t=-0.67\pm0.01$ for the total emission and of $\alpha_c=-0.62\pm0.01$ in the core region with Emission Measures of $EM_t=1.12\pm0.02\cdot 10^5\,\text{pc}\,\text{cm}^{-6}$ and $EM_c=3.16\pm0.1\cdot 10^5\,\text{pc}\,\text{cm}^{-6}$ respectively. Using the calculated values results in an opacity of the core region of $\tau_c=0.92$ at $\uplambda$92\,cm.
Since the Emission Measure is defined as
\begin{equation}
EM = \int_0^{s_0} n_e^2 ds,
\end{equation}
where $n_e$ is the electron density of the ionised medium and $s_0$ the line-of-sight through the medium, $n_e$ can easily be calculated. Using $s_0=900$\,pc results in an electron density of $n_e=18.7\,\text{cm}^{-3}$. Assuming that nearly all electrons are confined inside the star-forming regions and supernova remnants, the filling factor $f_e$ can be calculated. Using a density of $n_e=1000\,\text{cm}^{-3}$ from \citetads{2009ApJ...706.1571W} for the star-forming regions leads to a filling factor of $f_e=1.9\%$, which is confirmed by theoretical predictions from Youst-Hull et al. (priv. comm.).

These calculations suffer from a contribution to the integrated area of the disk emission in the front and on the backside of the galaxy as well as the unknown morphology in the core region. But anyhow they are good enough for an order of magnitude estimate.

\section{Magnetic field strength}
\label{text_magneticfield}
\label{text_total_magneticfield}

Assuming equipartition between the magnetic field energy and the energy density of the cosmic rays, one way to determine the total magnetic field strength is using the revised equipartition formula by \citetads{2005AN....326..414B}.

Due to the complicated structure of M82, the high dynamic range inside the galaxy in total intensity and the complex environment it is embedded in a pixel based method for the determination of the magnetic field strength and cosmic ray losses was used. This ensures that the steep gradient over the galaxy of the total power emission is accounted for and the error for the calculation of the magnetic field strength due to the integration over the line-of-sight is minimised. The $\uplambda22$\,cm fluxes were used for the entire calculation because of its superior dynamic range over the $\uplambda92$\,cm map and smaller influence of free-free absorption processes in the core region.

A cylindrical geometry was assumed for the synchrotron emitting regions, for which the projected profile is a Gaussian with constant values along the minor axis, and a Gaussian profile along the major axis. The full width half maximum for these geometries was calculated by convolving the 22\,cm map with the 92\,cm beam to match the resolution to the spectral index map making the result comparable, integrating the emission along the minor axis and fitting a Gaussian to the resulting profile yielding a FWHM of 930\,pc. Moreover, the assumed non-thermal spectral index was used from the halo fit in Fig. \ref{image_SI} since the equipartition between the magnetic field and cosmic ray energy is assumed to take several Myrs. This leads to reliable magnetic field strengths only in the halo region. To estimate the magnetic field strength in the core region a spectral index of $\upalpha=-0.6$ was used from the spectral index map between $\uplambda$6\,cm and $\uplambda$22\,cm (Fig. \ref{image_SI_6cm_22cm}), which is less affected in this region by ionisation losses. The corresponding fluxes and intensities were used from the $\uplambda$22\,cm map.

\begin{figure}
	\resizebox{\hsize}{!}{\includegraphics{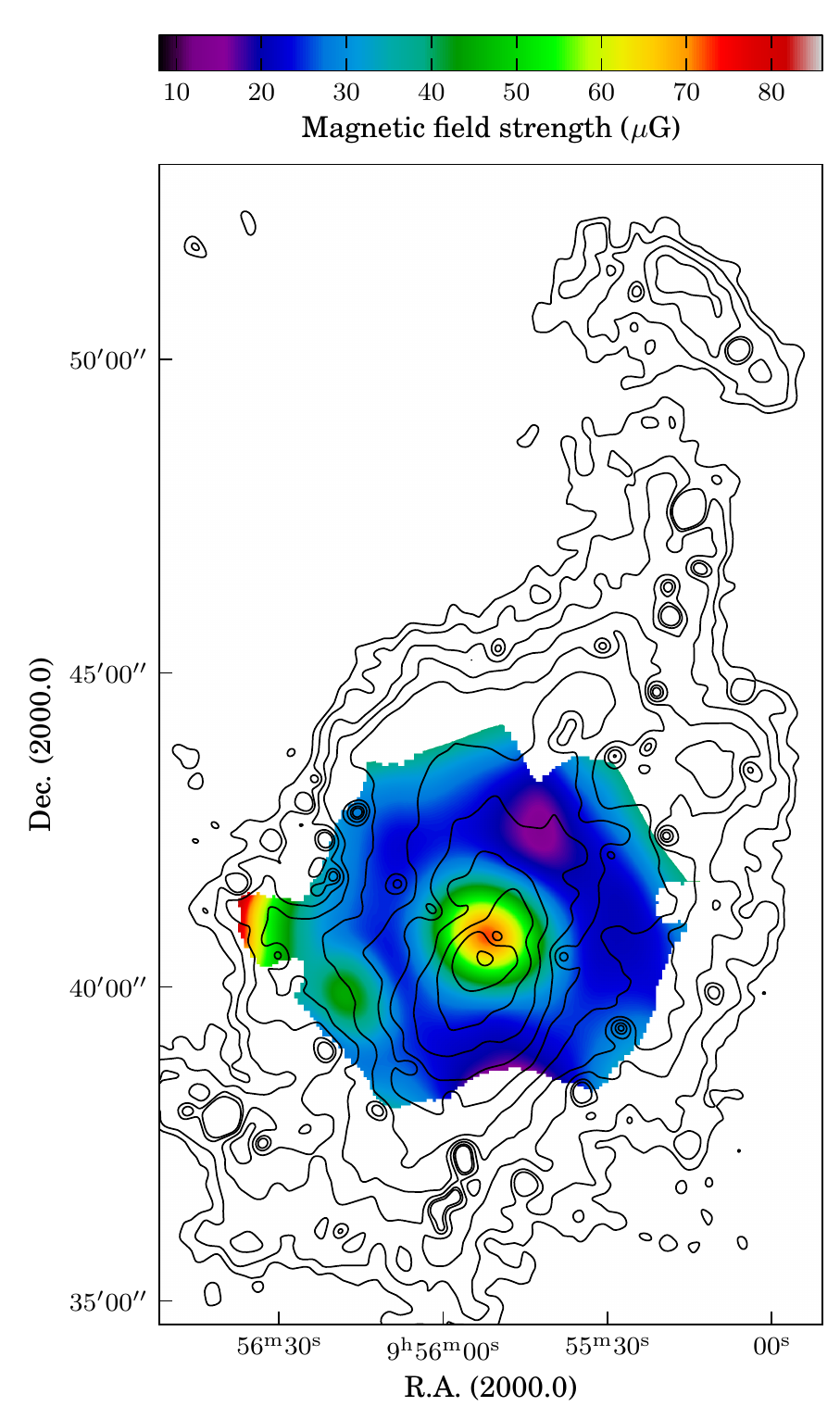}}
	\caption{Maps of the average equipartition strength of the magnetic field for the assumed cylindrical geometry of the synchrotron emitting medium using the intensities at $\uplambda22\,$cm and the spectral index in the halo from Fig. \ref{image_SI}. The overlaid are the soft X-ray (0.2-1\,keV) contours from available XMM-Newton archive data (We\.zgowiec et al. in prep.). The resolution of the image is 15\arcsec.}
	\label{image_bfield}
\end{figure}

The pixel based technique of calculating the magnetic field strength is only applicable for the total intensity emission as an estimate of the total magnetic field strength. The polarisation and its implication to the ordered magnetic field in M82 will be discussed in Paper II. The number density ratio of protons to electrons was assumed to be constant at a value of $K_0 = 100$. This assumption may not be fulfilled due to the possible high fraction of protons suffering less energy losses due to their higher mass than the electrons on their path away from the disk. In contrast to that the production of secondary electrons via pion decay would lead to a decrease of $K_0$. It is very likely that both effects are viable and could compensate each other. The uncertainty here only scales approximately with the fourth root of the assumed ratio, so that even for values between $K_0=10$ to $K_0=1000$ the error is less than $\pm30$\,\%.

The map resulting from these calculations is presented in Fig. \ref{image_bfield}. Averaging the magnetic field strength over the whole galaxy results in a mean equipartition field strength of $B_{eq}\approx35\,\upmu$G for the assumed geometry. This is high but still close to the values for spiral galaxies, which were calculated as having $9\pm3\,\upmu$G from a sample of 74 galaxies \citepads{1995A&AS..114...21N}. For galaxies with higher star formation rates like M51 the average field strengths are $20\,\upmu$G \citepads{2011MNRAS.412.2396F}. Looking into the central region of M82 gives an even higher magnetic field strength of $B_{eq}\approx98\,\upmu$G. This may be uncertain by an order of magnitude for the above mentioned reasons but is of the same order as the value derived by \citetads{1988A&A...190...41K} ($B_{eq}\approx50\,\upmu$G). Lacki \& Beck (in prep.) suggested a revised equipartition formula for starburst galaxies, which includes pion decay processes and the generation of secondary electrons. They calculated a magnetic field strength of $B_{eq}=250\,\upmu$G, but this value was derived by using the integrated flux over the whole galaxy and not distinguishing between the core and halo emission as well as using a constant scalelength of $l=500$\,pc and is therefore higher.

The calculated average halo field strength for M82 is $B_{eq}\approx24\,\upmu$G. Some rising gradient is visible towards the edges of the map. Since the resulting field strength is proportional to $(1/l)^\alpha$, the assumed geometry influences the values. A more spherical geometry would result in a more homogeneous magnetic field gradient over the halo. Since the geometry of the synchrotron emitting medium is mostly unknown, the errors are quite huge. But anyhow it has to be stressed that the cylindrical geometry seems to be the preferred one for M82 due to its correlation to other wavelengths like H$\upalpha$ \citepads{2002PASJ...54..891O} or X-ray \citepads{2009ApJ...697.2030S}. These H$\upalpha$ observations also showed two different components of the outflow cone, one with a small opening angle and a cylindrical geometry and the other with a larger opening angle and a plume-like geometry. A magnetic field morphology following the latter one would reduce the magnetic field strength in the mentioned regions.

\section{Cosmic ray electron losses}
\label{label_losses}

Cosmic ray electrons, which are assumed to be produced in star-forming regions mainly suffer energy losses from five different processes: Synchrotron radiation, the inverse Compton (IC) effect, non-thermal bremsstrahlung, ionisation, and adiabatic losses. In order to get an estimate of their contribution to the total losses, the different timescales were calculated for the derived values for the core and the halo region. This can be achieved using the equations given by \citetads{2011hea..book.....L}. All derived values are listed in Tab. \ref{table_crlosses}.

Electrons gyrating in a magnetic field $B\,(\upmu\text{G})$ of a given strength and radiating at a frequency $\nu\text{(Hz)}$ have the energy $E\text{(GeV)}$ according to the equation:
\begin{equation}
E \approx 2.5\cdot 10^{-4}\nu^\frac{1}{2} B^{-\frac{1}{2}}\,\text{GeV}
\end{equation}
The synchrotron lifetime $\tau_{syn}$ can then be calculated using:
\begin{equation}
\tau_{syn} \approx 8.35\cdot 10^9 E^{-1} B^{-2}\,\text{yrs,}
\end{equation}
To calculate the inverse Compton lifetime $\tau_{IC}$ an energy density $U_{ph}\,\left(\text{erg}\,\text{cm}^{-3}\right)$ is needed:
\begin{equation}
\tau_{IC} \approx 3.55\cdot 10^{-4} E^{-1} {U_{ph}}^{-1}\,\text{yrs,}
\end{equation}
This can be calculated using the Radio-FIR correlation from \citetads{1988ApJS...68..151H}
\begin{equation}
L_{FIR} = 1.26\cdot 10^{-14} \cdot (2.58\,S_{60} + S_{100})\,\text{erg cm}^{-3},
\end{equation}
with $L_{FIR}$ being the FIR luminosity in W\,m$^{-2}$ and $S_{60/100}$ the FIR luminosity in Jy at wavelengths of $60\,\upmu$m and $100\,\upmu$m. Using a value of $S_{60}=630$\,Jy and $S_{100}=529$\,Jy for the central $1\arcmin$ radius and $S_{60}=1700$\,Jy and $S_{100}=1840$\,Jy for a radius of $4\arcmin$ representing the whole galaxy \citepads{2010A&A...514A..14K} leads to an energy density of $U_{ph,c} = 4.43\cdot 10^{-11}\,\text{erg}\,\text{cm}^{-3}$ for the core and $U_{ph,h} = 5.57\cdot 10^{-12}\,\text{erg}\,\text{cm}^{-3}$ for the halo.

To estimate the non-thermal bremsstrahlung lifetime $\tau_{brems}$ and the ionisation lifetime $\tau_{ion}$, the ISM particle density $n\,(\text{cm}^{-3})$ is needed:
\begin{equation}
\tau_{brems} \approx 4\cdot 10^7 n^{-1}\,\text{yrs}
\end{equation}
\begin{equation}
\tau_{ion} \approx 9.5\cdot 10^6 n^{-1} E\,\text{yrs}
\end{equation}
Using $n_c=250\,\text{cm}^{-3}$ for the core region from \citetads{2001A&A...365..571W} gives a lower limit since \citetads{2010ApJ...722..668N} reported particle densities of up to several $10^4\,\text{cm}^{-3}$. This would reduce $\tau_{brems}$ and $\tau_{ion}$ drastically making them the dominant loss processes for the core region. For the halo the measured HI mass of $0.75\cdot 10^9 M_\odot$ from \citetads{1994Natur.372..530Y} is used. Assuming a spherical volume with a radius of 5\,kpc leads to a particle density in the halo of $n_h = 0.058\,\text{cm}^{-3}$.

The adiabatic losses are only dependent on the derivative of the velocity in direction of the minor axis:
\begin{equation}
\tau_{ad} = 3 \left(\frac{d\text{v}}{dz}\right)^{-1}\,\text{yrs}
\end{equation}
The values in the literature differ between velocities of $\text{v}=1200\,\text{km}\,\text{s}^{-1}$ \citepads{1992A&A...256...10R} for a non reaccelerated cosmic ray transport and $\text{v}=4000\,\text{km}\,\text{s}^{-1}$ \citepads{1991ApJ...369..320S} for one with reacceleration, but with the close correlation between the radio continuum morphology and the kinematic studies using charged particles, we can assume that the magnetic field is frozen into the ionised medium and moving at the same speed. Therefore we can take values of $\text{v}=600\,\text{km\,s}^{-1}$ at a height above the disk of $z=450$\,pc from \citetads{1988Natur.334...43B}, which is very close to the value derived for starbursting galaxies by \citetads{1999ApJ...513..156M} of $\text{v}=500\,\text{km}\,\text{s}^{-1}$. The cosmic ray bulk speed is an addition of the outflow speed of the charged particles and the Alfv\'en speed $\text{v}_A$. The latter one can be calculated using the standard textbook formula
\begin{equation}
\text{v}_A = 2.18\,n_e^{-\frac{1}{2}} B\,\text{km\,s}^{-1}.
\end{equation}
This results in a value of $\text{v}_A=10\,\text{km\,s}^{-1}$ using $n_e=30\,\text{cm}^{-3}$ from \citetads{1995A&A...293..703M} showing that in this environment the Alfv\'en speed is contributing negligibly to the bulk speed. Even using the value found in the cap region by \citetads{1999ApJ...510..197D} of $n_e=0.1\,\text{cm}^{-3}$ as a lower limit would result in an Alfv\'en speed of $\text{v}_A=165\,\text{km\,s}^{-1}$, which is still much lower than the outflow speed.

The adiabatic losses are still 1-2 orders lower than the ionisation and bremsstrahlung losses which are dominating the loss processes in the core region. This supports our interpretation of the lower fluxes in this region at longer wavelengths.

The \citetads{2009Natur.462..770V} detected $\upgamma$-ray emission from M82 with a luminosity of $L_{1\,GeV}\approx 1.9\cdot 10^{40}\,\text{erg\,s}^{-1}$ at 1\,GeV. Scaling this value down to 100\,MeV using their spectral index of $\alpha\approx 2.2$ results in $L_{100\,MeV}\approx 8.3\cdot10^{39}\,\text{erg\,s}^{-1}$. This value is of the same order as the predicted value of $L_{100\,MeV} = 3\cdot10^{39}\,\text{erg\,s}^{-1}$ using $L_{100\,MeV} = 2\cdot 10^{-5}L_{IR}$ \citepads{2006ApJ...645..186T} in case of pion decay. This would produce secondary electrons in the star-forming regions with the same spectrum as the protons and additionally flatten the spectrum towards lower frequencies. From the particle density $n$ one can calculate the pion loss timescale \citepads{1994A&A...286..983M}:
\begin{equation}
\tau_{\pi} \approx 2\cdot 10^5 \left(\frac{n}{250\,\text{cm}^{-3}}\right)^{-1}\,\text{yrs}
\label{pion_timescale}
\end{equation}
If this is shorter than $\tau_{esc}$ the starburst region is a proton calorimeter and pion decay has to be taken into account. This is the case for the calculated values. But care has to be taken here with the estimated values for $n$ since no filling factor was included in Eq. \ref{pion_timescale}. It is very unlikely that these secondary electrons are uniformly produced in the whole core region due to the high clumpiness of the neutral as well as ionised gas.

We can now calculate if the cosmic rays can escape the galaxy using the cooling timescale $\tau_{cool}$, which can be estimated for a continuous injection of cosmic rays by combining all the timescales:
\begin{equation}
\tau_{cool}^{-1} = \tau_{syn}^{-1} + \tau_{IC}^{-1} + \tau_{brems}^{-1} + \tau_{ion}^{-1} + \tau_{ad}^{-1} + \tau_{\pi}^{-1}
\end{equation}
This timescale can then be compared to the escape time for the cosmic rays due to advection with a galactic wind assuming a constant acceleration of the galactic wind from the core region into the halo:
\begin{equation}
\tau_{esc} \approx 1.02\,h\text{v}^{-1}\,\text{yrs},
\end{equation}
where $h$ is the scaleheight of the disk, which was calculated to $h=70\,$pc from the average of the first component of the two component fit to the scaleheights for the core region and to $h=670\,$pc from the exponential fit to the halo (Sect. \ref{text_scaleheights_22_92}), and v the bulk speed of the outflowing cosmic ray gas.
\begin{table}
	\caption{Overview of the timescales of the cosmic ray electrons. All values were calculated for the $\uplambda$22cm observations. The middle column represents the loss timescales for the core region and the right column the ones for the halo.}
	\label{table_crlosses}
	\centering
	\begin{tabular}{@{} lcc @{}}
		\toprule
		\toprule
		& $B_{core}=98\,\upmu$G & $B_{halo}=24\,\upmu$G \\
		\midrule
		$E$\,(GeV) & 0.94 & 1.90 \\
		$\tau_{syn}$\,(yrs) & $9.21\cdot 10^5$ & $7.62\cdot 10^6$ \\
		$\tau_{IC}$\,(yrs) & $8.51\cdot 10^6$ & $3.35\cdot 10^7$ \\
		$\tau_{brems}$\,(yrs) & $1.60\cdot 10^5$ & $6.90\cdot 10^8$ \\
		$\tau_{ion}$\,(yrs) & $3.57\cdot 10^4$ & $3.12\cdot 10^8$ \\
		$\tau_{ad}$\,(yrs) & $2.20\cdot 10^6$ & $2.20\cdot 10^6$ \\
		$\tau_{cool}$\,(yrs) & $2.44\cdot 10^4$ & $1.61\cdot 10^6$ \\
		$\tau_{esc}$\,(yrs) & $1.17\cdot 10^5$ & $1.12\cdot 10^6$ \\
		$\tau_{\pi}$\,(yrs) & $2.00\cdot 10^5$ & $8.62\cdot 10^8$ \\
		\bottomrule
	\end{tabular}
\end{table}

Comparing now the different timescales shows that synchrotron losses are not the dominant loss process for the cosmic ray electrons in the core region even assuming energy equipartition. Since $\tau_{cool}$ is an order of magnitude smaller than $\tau_{esc}$, the cosmic ray electrons cannot escape the recent starburst directly from the star bursting region into the halo within a normal galactic wind. Even more important is that $\tau_{cool}$ in the halo is of the same order as $\tau_{esc}$ making the build-up of a kiloparsec size halo difficult.

This shows that the existence of the large radio continuum halo around M82 is lacking an easy explanation and more complicated processes like secondary electrons originating from pion decay and multiple star-bursting episodes have to be taken into account.

\subsection{Building up the halo}

Using a simple hydrostatic equilibrium approach from \citetads{2006ApJ...645..186T} the magnetic field strength for the core region can be calculated with
\begin{equation}
B_{HE} = \sqrt{\frac{8\pi^2 \text{G}}{\eta f_g^{-1}}}\Sigma_g,
\end{equation}
where G is the gravitational constant, $\eta$ the ratio for which the gravitational and magnetic energy are in equilibrium, $f_g$ the complete gas fraction, and $\Sigma_g$ the surface density. Assuming a value of $\eta=1$ to set an upper limit for the magnetic field strength, a gas fraction of $f_g=\frac{1}{3}$ \citepads{2001ApJ...552..544F}, and a surface density of 0.85\,g\,cm$^{-2}$ \citepads{1998ApJ...498..541K} results in an average magnetic field strength of $B_{HE} = 1.13$\,mG. This would reduce $\tau_{syn}$ to $2.36\cdot 10^4$\,yrs and increase $\tau_{IC}$ to $2.89\cdot 10^7$\,yrs making synchrotron losses the dominant loss process in this region. Using these values and an average bulk speed for the cosmic rays of $\text{v$_c$}=1000\,\text{km\,s}^{-1}$ in the core region results in a travel distance of $s_c=24$\,pc within the loss timescale, which is too short to build up the observed extended synchrotron halo.

A possible explanation results from a scenario with a filling factor of the magnetic field similar to the ionised \ion{H}{ii} medium of $f_e=1.9\%$. Using our value of $B_{eq}=98\,\upmu$G for the core region results in an average magnetic field strength of $B\approx5$\,mG for the individual \ion{H}{ii}-regions remnants yielding a synchrotron loss timescale of $\tau_{syn}=2.42\cdot 10^3$\,yrs. Using the same value for the bulk speed as above gives a travel distance of $s_c=2.5$\,pc. \citetads{1994MNRAS.266..455M} found an average size of $2.4\pm1.0\,$pc for the compact sources in M82. This value is by more than one order of magnitude smaller than the one for compact sources in the Large Magellanic Cloud or the Milky Way and therefore magnetic field strength in the mG regime are a reasonable value. It was interpreted as being smaller due to its young population, but it can also be smaller due to the higher pressure of the medium surrounding those complexes and constraining their size. This would additionally support the arguments for a low filling factor of the ionised medium and the magnetic field.

\begin{figure}
	\resizebox{\hsize}{!}{\includegraphics{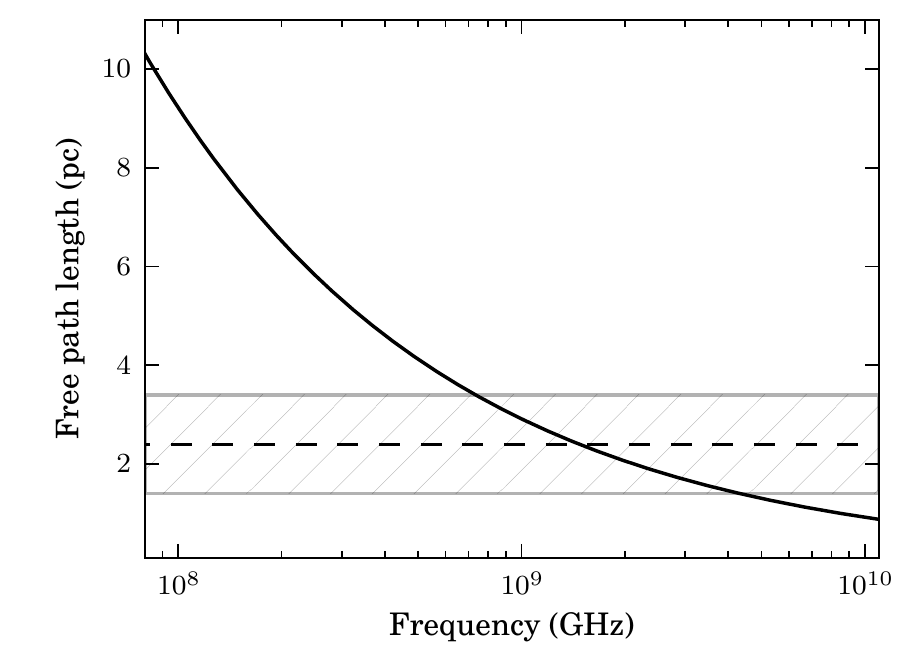}}
	\caption{Plot of the travel distance for the cosmic ray electrons vs. their emitted frequency inside the core region under the assumption of a magnetic field of $B\approx5$\,mG and a bulk speed of $\text{v$_c$}=1000\,\text{km\,s}^{-1}$. The straight line shows the average size for the compact sources from \citetads{1994MNRAS.266..455M} with their errors marked as shaded. The two lines cross each other at $\nu=1.4$\,GHz.}
	\label{plot_crlosses}
\end{figure}

Fig. \ref{plot_crlosses} shows the dependency of the cosmic ray energy vs. their free path lengths in case of an exploding supernova inside an \ion{H}{ii}-region using the above mentioned field strength and velocity at frequencies between 100\,MHz and 10\,GHz. A comparison with the average compact source size (dashed line) shows that only electrons injected at frequencies $\nu<1.4$\,GHz can escape the \ion{H}{ii}-regions. Since synchrotron emission from the halo is only visible at $\uplambda$22\,cm and $\uplambda$92\,cm this explanation describes the observations well.

The pressure to keep the compact regions confined would then originate from the hot X-ray gas as well as the molecular and neutral gas filling the cavity in between them with a filling factor close to 100\,\%. The gas density and photon pressure is up to $10^4$ times higher than those in the Milky Way on larger scales resulting in a much higher filling factor for these gas phases. A simple estimate using pressure equilibrium between the ionised gas in the supernova- and \ion{H}{ii}-regions with a temperature of $10^4$\,K and the X-ray gas with an assumed temperature of $10^7$\,K leads to an energy density of 50\,eV\,cm$^{-3}$. This is on the same order as the value derived by \citetads{1988A&A...190...41K} (60\,eV\,cm$^{-3}$), and very similar to our calculated photon field strength of 28\,eV\,cm$^{-3}$.

Since we showed that electrons produced by supernovae in contrast to their accompanying protons already lose a large part of their injected energy over the first parsecs of their travel, the ratio between the energy densities in protons and electrons increases towards the halo. Additionally, the hot ionised gas filling the whole galaxy is charged and will couple the gyrating electrons to its kinematics together with the magnetic field as soon as they leave the high magnetic field in the ionised star-forming complexes. This explains the strong correlation between the H$\upalpha$-, PAH$^+$-, and radio continuum morphologies and still accounts for the observed indications concerning pion decay and secondary electrons.

Using a simple estimate for the travel distance in the halo with a cosmic ray bulk speed of $\text{v}=600\,\text{km\,s}^{-1}$ and the cooling timescale of $\tau_{cool,h} = 1.61\cdot 10^6\,$yrs results in a travel distance of $s=990$\,pc. This is still by a factor of four lower than the observed 4\,kpc extent of the synchrotron halo. A radial decrease of the dust temperature in the halo directing outwards was observed by \citetads{2010A&A...518L..66R}, which indicates lower values for the radiation field. This would increase the travel distance by decreasing the inverse Compton losses, but not the adiabatic or synchrotron losses, which dominate the loss timescales in the halo and are limiting the travel distances to the estimated value. A dominance of synchrotron losses in the halo would additionally produce a steepening of the spectral index in this region, which is not observed. A scenario, which agrees with the observations, is the build-up of the halo within multiple starbursting periods, where the magnetic field and the bulk speed vary allowing the fuelling of the halo with fresh material.

\subsection{Multiple starbursting periods}

\citetads{2003ApJ...599..193F} describe a scenario of at least two successive starburst epochs for M82 about 10 and 5\,Myrs ago in addition to the recent one, where the material was released into the halo and has been accelerated by shocks and the high pressure from the core region \citepads{2010A&A...514A..14K}. This would also explain the observed kpc extent of the synchrotron halo. Within periods where the magnetic field due to less star-formation declines and adiabatic losses are less efficient due to a lower bulk speed, the cosmic rays can leave the inner region of the galaxy and build up a halo of the observed size. This would need variations of the loss timescales with at least a factor of four over the time of a starburst period to reach heights above the disk of several kpcs. \citetads{1996A&A...314..745L} showed that variations in the far-infrared and radio luminosity of one order of magnitude are reasonable for young, compact starbursts like M82.

Such an evolutionary scenario would not only explain the extent of the radio halo, but also the abundance of PAH$^+$ in the halo, which would if expelled into the halo from the current starburst have been destroyed due to the strong radiation from the radiation field \citepads{2008MNRAS.389..629B}. Furthermore, the hot gas cap visible far to the north in Fig. \ref{image_bfield} might be a remnant of an older starburst phase at the interface of the intergalactic medium.

The difference between the extents of the synchrotron halo at $\uplambda$22/$\uplambda$92\,cm and $\uplambda$3/$\uplambda$6\,cm is now explainable. The recently injected higher energetic cosmic rays would not reach the halo due to their high synchrotron losses in the compact \ion{H}{ii}-regions in the core. In this case the cosmic rays could be expelled into the halo due to their increased escape timescale, which is not possible during a starburst period like M82 is encountering now.

Possibilities for particle reacceleration and diffusion were discussed by \citetads{1991ApJ...369..320S}. While reacceleration by small-scale turbulence would lead to depolarisation inside the beam, large scale shocks originating from an epoch of starburst activity can still reaccelerate the electrons in the halo and lead to observable polarisation degrees as seen by \citetads{1994A&A...282..724R}. On the other hand the recent high-resolution H$\upalpha$-observations show small shocks inside the outflow cone suggesting that at least on these scales the cosmic ray electrons reside inside a small-scale turbulent field. This cannot rule out that on larger scales the outflow may still be polarised. \citetads{1980ApJ...239.1089L} suggested a diffusion dominated model which would result in a flat spectral index gradient for the halo. This would agree with the observations and therefore may be the preferred one. Since the analysis of the polarisation data is going to be discussed in Paper II, we refer to this for a detailed discussion.

\section{Summary}
\label{text_summary}

A collection of datasets from the VLA at $\uplambda3$\,cm and $\uplambda6$\,cm and from the WSRT at $\uplambda22$\,cm and $\uplambda92$\,cm was reduced to carry out a multifrequency analysis of the starburst galaxy M82. The physical parameters in the core region as well as its influence on the halo were investigated and connected to the evolution scenario of the galaxy.

Advanced data calibration techniques for radio continuum data are needed to achieve dynamic ranges sufficient for the scientific analysis of strong extended sources like M82. With the newly proposed technique in this publication we were able to reach dynamic ranges of $\sim50000$ for $\uplambda22$\,cm observations and even $\sim10000$ for $\uplambda92$\,cm.

These high dynamic range images allowed us to look into the morphology of the halo at these long wavelengths. While the extent of the halo at $\uplambda3$\,cm and $\uplambda6$\,cm is limited to the inner 1\,kpc radius from the core, diffuse emission up to 4\,kpc radius shows up at $\uplambda22$\,cm and $\uplambda92$\,cm. A large difference in the extent of the northern and southern outflow is recognised. While the northern emission reaches up to 4-5\,kpc into the halo, the southern one can only be traced up to 2.5\,kpc. This asymmetry is also visible in the X-ray observations (Fig. \ref{image_bfield}). A possible explanation would be a motion of M82 towards the group centre in the south causing ram pressure effects like in clusters of galaxies, but such an evolutionary scenario contradicts the results by \citetads{1993egte.conf..253Y}.

The northern outflow cone shows a lack of emission at long wavelengths in contrast to the two spurs surrounding it and steep spectral indices of up to $-2.5$ in its central part. This indicates that the dominant loss processes for the outflow cone in the north are either adiabatic or synchrotron losses. Numerical studies by \citetads{2007Ap&SS.309..151M} confirm this in the case of NGC253, where a similar but not as well pronounced morphology was recognized by \citetads{2009A&A...494..563H}.

A fitting to the scaleheights lead to very different results than in spiral galaxies. While for spiral galaxies typically a thin and a thick disk is found with scaleheights of $0.3\,$kpc and $1.8\,$kpc respectively \citepads{2011arXiv1111.7081K}, the values for M82 were lower by a factor of $\sim3$. Additionally an overall small gradient for the spectral index was found for the halo between $\uplambda22$\,cm and $\uplambda92$\,cm with $\alpha=-1.2--1.5$, while an increase is expected for galaxies where inverse Compton and synchrotron losses dominate the cosmic ray losses. Using an energy equipartition ansatz between the magnetic field energy and the cosmic ray particles results in an average field strength of $24\,\upmu$G for the halo and $98\,\upmu$G for the core region.

An integration over the emission of the core and the halo showed that free-free absorption is lowering the flux of the core at wavelengths from $\uplambda22\,$cm on, while the halo emission is completely free of this feature at least up to $\uplambda92$\,cm indicating a high contribution of bremsstrahlung losses in the core region. From a fit to the data of the core an average electron density of $19\,\text{cm}^{-3}$ could be obtained leading to a filling factor of $\sim1.9\%$ for the ionised hydrogen. These fits are limited to just five datapoints and therefore have to be taken with care, but additional observations between $\uplambda22$\,cm and $\uplambda92$\,cm with the WSRT and even at lower wavelengths with LOFAR will give better constrains on this.

A calculation of the loss timescales shows that at the current stage the ionisation and bremsstrahlung losses are dominating the expansion of the cosmic rays in the core region confirming the previous statement. Additionally the recent $\upgamma$-ray detections support the thesis of M82 encountering pion decay in the core region indicating very high particle densities, but lack the resolution to determine the exact origin of these particles within the galaxy.

To account for the above observations we pointed out a scenario where \ion{H}{ii}-regions and supernova remnants are dominating the radio flux in the core region. Cosmic rays are generated inside these small regions, where the medium is so dense that secondary electron production via pion decay is possible and cosmic ray electrons are suffering high losses due to strong magnetic fields. Using the same filling factor for the magnetic field as for the ionised medium, the magnetic field strength inside these compact regions was estimated to $5.16$\,mG. This makes synchrotron losses the dominant loss process inside these regions and the free path lengths for these particles at $\nu\geq1.4\,$GHz are smaller than the size of these regions. This explains the difference of the extent of the synchrotron halo between frequencies of $\uplambda$3\,cm/$\uplambda$6\,cm and $\uplambda$22\,cm/$\uplambda$92\,cm. Higher energetic particles lose their energy faster than lower energetic ones and are not propagated out into the halo. Since the pion decay mostly occurs inside these dense regions, it cannot help to produce electrons less affected by these high synchrotron losses.

In contrast to this heavier charged particles like protons or molecules are less affected by the magnetic field and therefore can escape these small regions more easily. The pressure to constrain these regions to sizes of several pc can be generated by the hot ionised X-ray and the neutral hydrogen and molecular gas filling the whole core region of the galaxy. As soon as charged particles reach the lower density regions surrounding the star-forming regions they become coupled to the kinematic flow of the lower density X-ray gas and may leave the core region, which explains the similarity in the morphologies of H$\upalpha$, PAH$^+$ and radio.

The loss timescale in the halo is of the same order as the escape timescale. Using a simple scenario, where cosmic rays are propagated from the core region into the halo results in free path lengths of only 990\,pc, which is too low to build up a 4\,kpc size halo. This gives evidence for a scenario where M82 experienced several starburst periods in the last 5-10\,Myrs, in which the FIR- and radio-luminosity varied by one order of magnitude and it was possible to eject cosmic rays into the halo. The small gradient in the spectral index in the halo indicates that different propagation effects than in normal spiral galaxies are dominant. It cannot explain a halo where a continuous propagation of cosmic rays suffering from inverse Compton and synchrotron losses is fuelling the halo. Instead we propose that the synchrotron halo was produced by multiple ejections of cosmic rays and the observed spectrum is a superposition of several populations of those.

\section{Conclusions}
\label{text_conclusions}

The described scenario could imply that the magnetisation of the intergalactic medium by starbursts is limited to intervals shorter than the actual time of the starburst since losses for the cosmic rays are too high to let cosmic ray electrons escape the starbursting galaxy as soon as a strong magnetic field and high densities have been reached. Only heavier charged particles might still escape the galaxy due to their longer loss timescales. This would have a significant influence on the proton to electron ratio in the intergalactic medium and significantly influence theoretical predictions of the propagation of cosmic rays and the strength of magnetic fields in the intergalactic space. For the first time \citetads{2011MNRAS.415.3189K} showed in simulations of three interacting galaxies, that the whole group medium can become magnetised over a timescale of several hundred Myrs.

This evolution scenario and the discovered high clumpiness of the medium in M82 even has influence on predictions for the early Universe. Due to the overall increased star-formation rate and number of starburst galaxies the release of magnetic fields into the intergalactic medium may not be as easy as predicted by current simulations. Additionally, the high pressure environments in such galaxies might sufficiently suppress the expansion of supernova shells. Hence the dominant mechanism for material leaving the originating region are large scale shocks from only several supernova explosions inside an \ion{H}{ii}-region. To verify our evolution scenario further numerical MHD-simulations are needed using the discussed loss processes together with the constrains given here and in several other publications.

With LOFAR becoming more and more scientifically usable and complementary observations from the WSRT and VLA with wavelength of up to $\uplambda92$\,cm for a lot of nearby galaxies the free-free absorption at long wavelengths is a viable technique to estimate the thermal electron density and the filling factor of the ionised medium of star-forming spiral and starbursting galaxies. The integrated spectrum over whole galaxies does not sufficiently account for the difference between the complicated physics in the star-forming regions and the halo and interarm areas. Due to their high resolution the mentioned observations are able to differentiate between these regions and provide new input for theoretical predictions and numerical simulations.

\acknowledgements{The authors wish to express thanks to Rainer Beck, John S. Gallagher, and V. Heesen for useful scientific discussions and to Mark Westmoquette for providing the H$\upalpha$ map and additional input to the publication. We want to express our gratitude to Carlos Sotomayor from the Astronomisches Institut der Ruhr-Universit\"at Bochum for his help in developing scripts for the WSRT data reduction. The work at the AIRUB is supported by the DFG through FOR 1048. The Westerbork Synthesis Radio Telescope is operated by ASTRON (Netherlands Foundation for Research in Astronomy) with support from the Netherlands Foundation for Scientific Research (NWO). The National Radio Astronomy Observatory is a facility of the National Science Foundation operated under cooperative agreement by Associated Universities, Inc.}

\bibliography{/home/bjorn/Documents/BibTeX/bibtex}{}

\begin{thebibliography}{65}
\expandafter\ifx\csname natexlab\endcsname\relax\def\natexlab#1{#1}\fi

\bibitem[{{Aladro} {et~al.}(2011){Aladro}, {Mart{\'{\i}}n},
  {Mart{\'{\i}}n-Pintado}, {Mauersberger}, {Henkel}, {Oca{\~n}a Flaquer}, \&
  {Amo-Baladr{\'o}n}}]{2011A&A...535A..84A}
{Aladro}, R., {Mart{\'{\i}}n}, S., {Mart{\'{\i}}n-Pintado}, J., {et~al.} 2011,
  \aap, 535, A84

\bibitem[{{Beck} {et~al.}(1996){Beck}, {Brandenburg}, {Moss}, {Shukurov}, \&
  {Sokoloff}}]{1996ARA&A..34..155B}
{Beck}, R., {Brandenburg}, A., {Moss}, D., {Shukurov}, A., \& {Sokoloff}, D.
  1996, \araa, 34, 155

\bibitem[{{Beck} \& {Krause}(2005)}]{2005AN....326..414B}
{Beck}, R. \& {Krause}, M. 2005, Astronomische Nachrichten, 326, 414

\bibitem[{{Bendo} {et~al.}(2008){Bendo}, {Draine}, {Engelbracht}, {Helou},
  {Thornley}, {Bot}, {Buckalew}, {Calzetti}, {Dale}, {Hollenbach}, {Li}, \&
  {Moustakas}}]{2008MNRAS.389..629B}
{Bendo}, G.~J., {Draine}, B.~T., {Engelbracht}, C.~W., {et~al.} 2008, \mnras,
  389, 629

\bibitem[{{Bland} \& {Tully}(1988)}]{1988Natur.334...43B}
{Bland}, J. \& {Tully}, B. 1988, \nat, 334, 43

\bibitem[{{Braun} {et~al.}(2007){Braun}, {Oosterloo}, {Morganti}, {Klein}, \&
  {Beck}}]{2007A&A...461..455B}
{Braun}, R., {Oosterloo}, T.~A., {Morganti}, R., {Klein}, U., \& {Beck}, R.
  2007, \aap, 461, 455

\bibitem[{{Brentjens}(2008)}]{2008A&A...489...69B}
{Brentjens}, M.~A. 2008, \aap, 489, 69

\bibitem[{{Chynoweth} {et~al.}(2008){Chynoweth}, {Langston}, {Yun}, {Lockman},
  {Rubin}, \& {Scoles}}]{2008AJ....135.1983C}
{Chynoweth}, K.~M., {Langston}, G.~I., {Yun}, M.~S., {et~al.} 2008, \aj, 135,
  1983

\bibitem[{{Clark}(1980)}]{1980A&A....89..377C}
{Clark}, B.~G. 1980, \aap, 89, 377

\bibitem[{{de Bruyn} \& {Brentjens}(2005)}]{2005A&A...441..931D}
{de Bruyn}, A.~G. \& {Brentjens}, M.~A. 2005, \aap, 441, 931

\bibitem[{{de Mello} {et~al.}(2008){de Mello}, {Smith}, {Sabbi}, {Gallagher},
  {Mountain}, \& {Harbeck}}]{2008AJ....135..548D}
{de Mello}, D.~F., {Smith}, L.~J., {Sabbi}, E., {et~al.} 2008, \aj, 135, 548

\bibitem[{{Devine} \& {Bally}(1999)}]{1999ApJ...510..197D}
{Devine}, D. \& {Bally}, J. 1999, \apj, 510, 197

\bibitem[{{Dumke} \& {Krause}(1998)}]{1998LNP...506..555D}
{Dumke}, M. \& {Krause}, M. 1998, in Lecture Notes in Physics, Berlin Springer
  Verlag, Vol. 506, IAU Colloq. 166: The Local Bubble and Beyond, ed.
  {D.~Breitschwerdt, M.~J.~Freyberg, \& J.~Truemper}, 555--558

\bibitem[{{Dumke} {et~al.}(1995){Dumke}, {Krause}, {Wielebinski}, \&
  {Klein}}]{1995A&A...302..691D}
{Dumke}, M., {Krause}, M., {Wielebinski}, R., \& {Klein}, U. 1995, \aap, 302,
  691

\bibitem[{{Fenech} {et~al.}(2010){Fenech}, {Beswick}, {Muxlow}, {Pedlar}, \&
  {Argo}}]{2010MNRAS.408..607F}
{Fenech}, D., {Beswick}, R., {Muxlow}, T.~W.~B., {Pedlar}, A., \& {Argo}, M.~K.
  2010, \mnras, 408, 607

\bibitem[{{Fletcher} {et~al.}(2011){Fletcher}, {Beck}, {Shukurov},
  {Berkhuijsen}, \& {Horellou}}]{2011MNRAS.412.2396F}
{Fletcher}, A., {Beck}, R., {Shukurov}, A., {Berkhuijsen}, E.~M., \&
  {Horellou}, C. 2011, \mnras, 412, 2396

\bibitem[{{F{\"o}rster Schreiber} {et~al.}(2001){F{\"o}rster Schreiber},
  {Genzel}, {Lutz}, {Kunze}, \& {Sternberg}}]{2001ApJ...552..544F}
{F{\"o}rster Schreiber}, N.~M., {Genzel}, R., {Lutz}, D., {Kunze}, D., \&
  {Sternberg}, A. 2001, \apj, 552, 544

\bibitem[{{F{\"o}rster Schreiber} {et~al.}(2003){F{\"o}rster Schreiber},
  {Genzel}, {Lutz}, \& {Sternberg}}]{2003ApJ...599..193F}
{F{\"o}rster Schreiber}, N.~M., {Genzel}, R., {Lutz}, D., \& {Sternberg}, A.
  2003, \apj, 599, 193

\bibitem[{{Greaves} {et~al.}(2000){Greaves}, {Holland}, {Jenness}, \&
  {Hawarden}}]{2000Natur.404..732G}
{Greaves}, J.~S., {Holland}, W.~S., {Jenness}, T., \& {Hawarden}, T.~G. 2000,
  \nat, 404, 732

\bibitem[{{Gregory} \& {Condon}(1991)}]{1991ApJS...75.1011G}
{Gregory}, P.~C. \& {Condon}, J.~J. 1991, \apjs, 75, 1011

\bibitem[{{Heesen} {et~al.}(2009){Heesen}, {Beck}, {Krause}, \&
  {Dettmar}}]{2009A&A...494..563H}
{Heesen}, V., {Beck}, R., {Krause}, M., \& {Dettmar}, R.-J. 2009, \aap, 494,
  563

\bibitem[{{Helou} {et~al.}(1988){Helou}, {Khan}, {Malek}, \&
  {Boehmer}}]{1988ApJS...68..151H}
{Helou}, G., {Khan}, I.~R., {Malek}, L., \& {Boehmer}, L. 1988, \apjs, 68, 151

\bibitem[{{Hoopes} {et~al.}(2005){Hoopes}, {Heckman}, {Strickland}, {Seibert},
  {Madore}, {Rich}, {Bianchi}, {Gil de Paz}, {Burgarella}, {Thilker},
  {Friedman}, {Barlow}, {Byun}, {Donas}, {Forster}, {Jelinsky}, {Lee},
  {Malina}, {Martin}, {Milliard}, {Morrissey}, {Neff}, {Schiminovich},
  {Siegmund}, {Small}, {Szalay}, {Welsh}, \& {Wyder}}]{2005ApJ...619L..99H}
{Hoopes}, C.~G., {Heckman}, T.~M., {Strickland}, D.~K., {et~al.} 2005, \apjl,
  619, L99

\bibitem[{{Huchtmeier} \& {Skillman}(1998)}]{1998A&AS..127..269H}
{Huchtmeier}, W.~K. \& {Skillman}, E.~D. 1998, \aaps, 127, 269

\bibitem[{{Jacobs} {et~al.}(2009){Jacobs}, {Rizzi}, {Tully}, {Shaya},
  {Makarov}, \& {Makarova}}]{2009AJ....138..332J}
{Jacobs}, B.~A., {Rizzi}, L., {Tully}, R.~B., {et~al.} 2009, \aj, 138, 332

\bibitem[{{Kaneda} {et~al.}(2010){Kaneda}, {Ishihara}, {Suzuki}, {Ikeda},
  {Onaka}, {Yamagishi}, {Ohyama}, {Wada}, \& {Yasuda}}]{2010A&A...514A..14K}
{Kaneda}, H., {Ishihara}, D., {Suzuki}, T., {et~al.} 2010, \aap, 514, A14

\bibitem[{{Kennicutt}(1998)}]{1998ApJ...498..541K}
{Kennicutt}, Jr., R.~C. 1998, \apj, 498, 541

\bibitem[{{Klein} {et~al.}(1988){Klein}, {Wielebinski}, \&
  {Morsi}}]{1988A&A...190...41K}
{Klein}, U., {Wielebinski}, R., \& {Morsi}, H.~W. 1988, \aap, 190, 41

\bibitem[{{Kotarba} {et~al.}(2011){Kotarba}, {Lesch}, {Dolag}, {Naab},
  {Johansson}, {Donnert}, \& {Stasyszyn}}]{2011MNRAS.415.3189K}
{Kotarba}, H., {Lesch}, H., {Dolag}, K., {et~al.} 2011, \mnras, 415, 3189

\bibitem[{{Krause}(2009)}]{2009RMxAC..36...25K}
{Krause}, M. 2009, in Revista Mexicana de Astronomia y Astrofisica Conference
  Series, Vol.~36, Revista Mexicana de Astronomia y Astrofisica Conference
  Series, 25--29

\bibitem[{{Krause}(2011)}]{2011arXiv1111.7081K}
{Krause}, M. 2011, ArXiv e-prints 1111.7081

\bibitem[{{Kronberg} {et~al.}(1999){Kronberg}, {Lesch}, \&
  {Hopp}}]{1999ApJ...511...56K}
{Kronberg}, P.~P., {Lesch}, H., \& {Hopp}, U. 1999, \apj, 511, 56

\bibitem[{{Lacki} {et~al.}(2011){Lacki}, {Thompson}, {Quataert}, {Loeb}, \&
  {Waxman}}]{2011ApJ...734..107L}
{Lacki}, B.~C., {Thompson}, T.~A., {Quataert}, E., {Loeb}, A., \& {Waxman}, E.
  2011, \apj, 734, 107

\bibitem[{{Leeuw} \& {Robson}(2009)}]{2009AJ....137..517L}
{Leeuw}, L.~L. \& {Robson}, E.~I. 2009, \aj, 137, 517

\bibitem[{{Lehnert} {et~al.}(1999){Lehnert}, {Heckman}, \&
  {Weaver}}]{1999ApJ...523..575L}
{Lehnert}, M.~D., {Heckman}, T.~M., \& {Weaver}, K.~A. 1999, \apj, 523, 575

\bibitem[{{Lerche} \& {Schlickeiser}(1980)}]{1980ApJ...239.1089L}
{Lerche}, I. \& {Schlickeiser}, R. 1980, \apj, 239, 1089

\bibitem[{{Lisenfeld} {et~al.}(1996){Lisenfeld}, {Voelk}, \&
  {Xu}}]{1996A&A...314..745L}
{Lisenfeld}, U., {Voelk}, H.~J., \& {Xu}, C. 1996, \aap, 314, 745

\bibitem[{{Longair}(2011)}]{2011hea..book.....L}
{Longair}, M.~S. 2011, {High Energy Astrophysics, Cambridge University Press,
  3rd Edition, chapter 19}

\bibitem[{{Mannheim} \& {Schlickeiser}(1994)}]{1994A&A...286..983M}
{Mannheim}, K. \& {Schlickeiser}, R. 1994, \aap, 286, 983

\bibitem[{{Martin}(1999)}]{1999ApJ...513..156M}
{Martin}, C.~L. 1999, \apj, 513, 156

\bibitem[{{McDonald} {et~al.}(2002){McDonald}, {Muxlow}, {Wills}, {Pedlar}, \&
  {Beswick}}]{2002MNRAS.334..912M}
{McDonald}, A.~R., {Muxlow}, T.~W.~B., {Wills}, K.~A., {Pedlar}, A., \&
  {Beswick}, R.~J. 2002, \mnras, 334, 912

\bibitem[{{McKeith} {et~al.}(1995){McKeith}, {Greve}, {Downes}, \&
  {Prada}}]{1995A&A...293..703M}
{McKeith}, C.~D., {Greve}, A., {Downes}, D., \& {Prada}, F. 1995, \aap, 293,
  703

\bibitem[{{Miyake} {et~al.}(2007){Miyake}, {Yanagita}, \&
  {Yoshida}}]{2007Ap&SS.309..151M}
{Miyake}, S., {Yanagita}, S., \& {Yoshida}, T. 2007, \apss, 309, 151

\bibitem[{{Muxlow} {et~al.}(1994){Muxlow}, {Pedlar}, {Wilkinson}, {Axon},
  {Sanders}, \& {de Bruyn}}]{1994MNRAS.266..455M}
{Muxlow}, T.~W.~B., {Pedlar}, A., {Wilkinson}, P.~N., {et~al.} 1994, \mnras,
  266, 455

\bibitem[{{Naylor} {et~al.}(2010){Naylor}, {Bradford}, {Aguirre}, {Bock},
  {Earle}, {Glenn}, {Inami}, {Kamenetzky}, {Maloney}, {Matsuhara}, {Nguyen}, \&
  {Zmuidzinas}}]{2010ApJ...722..668N}
{Naylor}, B.~J., {Bradford}, C.~M., {Aguirre}, J.~E., {et~al.} 2010, \apj, 722,
  668

\bibitem[{{Niklas} {et~al.}(1995){Niklas}, {Klein}, {Braine}, \&
  {Wielebinski}}]{1995A&AS..114...21N}
{Niklas}, S., {Klein}, U., {Braine}, J., \& {Wielebinski}, R. 1995, \aaps, 114,
  21

\bibitem[{{Offringa} {et~al.}(2010){Offringa}, {de Bruyn}, {Biehl}, {Zaroubi},
  {Bernardi}, \& {Pandey}}]{2010MNRAS.405..155O}
{Offringa}, A.~R., {de Bruyn}, A.~G., {Biehl}, M., {et~al.} 2010, \mnras, 405,
  155

\bibitem[{{Ohyama} {et~al.}(2002){Ohyama}, {Taniguchi}, {Iye}, {Yoshida},
  {Sekiguchi}, {Takata}, {Saito}, {Kawabata}, {Kashikawa}, {Aoki}, {Sasaki},
  {Kosugi}, {Okita}, {Shimizu}, {Inata}, {Ebizuka}, {Ozawa}, {Yadoumaru},
  {Taguchi}, \& {Asai}}]{2002PASJ...54..891O}
{Ohyama}, Y., {Taniguchi}, Y., {Iye}, M., {et~al.} 2002, \pasj, 54, 891

\bibitem[{{Reuter} {et~al.}(1992){Reuter}, {Klein}, {Lesch}, {Wielebinski}, \&
  {Kronberg}}]{1992A&A...256...10R}
{Reuter}, H.-P., {Klein}, U., {Lesch}, H., {Wielebinski}, R., \& {Kronberg},
  P.~P. 1992, \aap, 256, 10

\bibitem[{{Reuter} {et~al.}(1994){Reuter}, {Klein}, {Lesch}, {Wielebinski}, \&
  {Kronberg}}]{1994A&A...282..724R}
{Reuter}, H.-P., {Klein}, U., {Lesch}, H., {Wielebinski}, R., \& {Kronberg},
  P.~P. 1994, \aap, 282, 724

\bibitem[{{Roussel} {et~al.}(2010){Roussel}, {Wilson}, {Vigroux}, {Isaak},
  {Sauvage}, {Madden}, {Auld}, {Baes}, {Barlow}, {Bendo}, {Bock}, {Boselli},
  {Bradford}, {Buat}, {Castro-Rodriguez}, {Chanial}, {Charlot}, {Ciesla},
  {Clements}, {Cooray}, {Cormier}, {Cortese}, {Davies}, {Dwek}, {Eales},
  {Elbaz}, {Galametz}, {Galliano}, {Gear}, {Glenn}, {Gomez}, {Griffin}, {Hony},
  {Levenson}, {Lu}, {O'Halloran}, {Okumura}, {Oliver}, {Page}, {Panuzzo},
  {Papageorgiou}, {Parkin}, {Perez-Fournon}, {Pohlen}, {Rangwala}, {Rigby},
  {Rykala}, {Sacchi}, {Schulz}, {Schirm}, {Smith}, {Spinoglio}, {Stevens},
  {Srinivasan}, {Symeonidis}, {Trichas}, {Vaccari}, {Wozniak}, {Wright}, \&
  {Zeilinger}}]{2010A&A...518L..66R}
{Roussel}, H., {Wilson}, C.~D., {Vigroux}, L., {et~al.} 2010, \aap, 518, L66

\bibitem[{{Sault} \& {Wieringa}(1994)}]{1994A&AS..108..585S}
{Sault}, R.~J. \& {Wieringa}, M.~H. 1994, \aaps, 108, 585

\bibitem[{{Seaquist} \& {Odegard}(1991)}]{1991ApJ...369..320S}
{Seaquist}, E.~R. \& {Odegard}, N. 1991, \apj, 369, 320

\bibitem[{{Soida} {et~al.}(2011){Soida}, {Krause}, {Dettmar}, \&
  {Urbanik}}]{2011A&A...531A.127S}
{Soida}, M., {Krause}, M., {Dettmar}, R.-J., \& {Urbanik}, M. 2011, \aap, 531,
  A127

\bibitem[{{Strickland} \& {Heckman}(2009)}]{2009ApJ...697.2030S}
{Strickland}, D.~K. \& {Heckman}, T.~M. 2009, \apj, 697, 2030

\bibitem[{{Thompson} {et~al.}(2006){Thompson}, {Quataert}, {Waxman}, {Murray},
  \& {Martin}}]{2006ApJ...645..186T}
{Thompson}, T.~A., {Quataert}, E., {Waxman}, E., {Murray}, N., \& {Martin},
  C.~L. 2006, \apj, 645, 186

\bibitem[{{Veilleux} {et~al.}(2009){Veilleux}, {Rupke}, \&
  {Swaters}}]{2009ApJ...700L.149V}
{Veilleux}, S., {Rupke}, D.~S.~N., \& {Swaters}, R. 2009, \apjl, 700, L149

\bibitem[{{VERITAS Collaboration} {et~al.}(2009){VERITAS Collaboration},
  {Acciari}, {Aliu}, {Arlen}, {Aune}, {Bautista}, {Beilicke}, {Benbow},
  {Boltuch}, {Bradbury}, {Buckley}, {Bugaev}, {Byrum}, {Cannon}, {Celik},
  {Cesarini}, {Chow}, {Ciupik}, {Cogan}, {Colin}, {Cui}, {Dickherber}, {Duke},
  {Fegan}, {Finley}, {Finnegan}, {Fortin}, {Fortson}, {Furniss}, {Galante},
  {Gall}, {Gibbs}, {Gillanders}, {Godambe}, {Grube}, {Guenette}, {Gyuk},
  {Hanna}, {Holder}, {Horan}, {Hui}, {Humensky}, {Imran}, {Kaaret}, {Karlsson},
  {Kertzman}, {Kieda}, {Kildea}, {Konopelko}, {Krawczynski}, {Krennrich},
  {Lang}, {Lebohec}, {Maier}, {McArthur}, {McCann}, {McCutcheon}, {Millis},
  {Moriarty}, {Mukherjee}, {Nagai}, {Ong}, {Otte}, {Pandel}, {Perkins},
  {Pizlo}, {Pohl}, {Quinn}, {Ragan}, {Reyes}, {Reynolds}, {Roache}, {Rose},
  {Schroedter}, {Sembroski}, {Smith}, {Steele}, {Swordy}, {Theiling},
  {Thibadeau}, {Varlotta}, {Vassiliev}, {Vincent}, {Wagner}, {Wakely}, {Ward},
  {Weekes}, {Weinstein}, {Weisgarber}, {Williams}, {Wissel}, {Wood}, \&
  {Zitzer}}]{2009Natur.462..770V}
{VERITAS Collaboration}, {Acciari}, V.~A., {Aliu}, E., {et~al.} 2009, \nat,
  462, 770

\bibitem[{{Walter} {et~al.}(2002){Walter}, {Weiss}, \&
  {Scoville}}]{2002ApJ...580L..21W}
{Walter}, F., {Weiss}, A., \& {Scoville}, N. 2002, \apjl, 580, L21

\bibitem[{{Wei{\ss}} {et~al.}(2001){Wei{\ss}}, {Neininger}, {H{\"u}ttemeister},
  \& {Klein}}]{2001A&A...365..571W}
{Wei{\ss}}, A., {Neininger}, N., {H{\"u}ttemeister}, S., \& {Klein}, U. 2001,
  \aap, 365, 571

\bibitem[{{Westmoquette} {et~al.}(2009){Westmoquette}, {Gallagher}, {Smith},
  {Trancho}, {Bastian}, \& {Konstantopoulos}}]{2009ApJ...706.1571W}
{Westmoquette}, M.~S., {Gallagher}, J.~S., {Smith}, L.~J., {et~al.} 2009, \apj,
  706, 1571

\bibitem[{{Wills} {et~al.}(1997){Wills}, {Pedlar}, {Muxlow}, \&
  {Wilkinson}}]{1997MNRAS.291..517W}
{Wills}, K.~A., {Pedlar}, A., {Muxlow}, T.~W.~B., \& {Wilkinson}, P.~N. 1997,
  \mnras, 291, 517

\bibitem[{{Yun} {et~al.}(1993{\natexlab{a}}){Yun}, {Ho}, {Brouillet}, \&
  {Lo}}]{1993egte.conf..253Y}
{Yun}, M.~S., {Ho}, P.~T.~P., {Brouillet}, N., \& {Lo}, K.~Y.
  1993{\natexlab{a}}, in Evolution of Galaxies and their Environment, ed.
  {J.~M.~Shull \& H.~A.~Thronson}, 253--254

\bibitem[{{Yun} {et~al.}(1993{\natexlab{b}}){Yun}, {Ho}, \&
  {Lo}}]{1993ApJ...411L..17Y}
{Yun}, M.~S., {Ho}, P.~T.~P., \& {Lo}, K.~Y. 1993{\natexlab{b}}, \apjl, 411,
  L17

\bibitem[{{Yun} {et~al.}(1994){Yun}, {Ho}, \& {Lo}}]{1994Natur.372..530Y}
{Yun}, M.~S., {Ho}, P.~T.~P., \& {Lo}, K.~Y. 1994, \nat, 372, 530

\end{thebibliography}
\bibliographystyle{aa}

\listofobjects

\end{document}